\documentclass[a4paper, 11pt]{article}
\usepackage[margin=25mm]{geometry}
\usepackage{booktabs}
\usepackage{multirow}
\usepackage{footnote}
\usepackage[shortlabels]{enumitem}
\usepackage{float}
\usepackage{comment}
\usepackage{amsmath, amssymb}
\usepackage{amsfonts}
\usepackage{titlesec}
\usepackage{bm}
\usepackage{lscape} 
\usepackage{graphicx,comment}
\usepackage{url}
\usepackage{color}
\usepackage{xargs, xstring}
\usepackage{verbatim}
\usepackage{setspace} 
\usepackage[round]{natbib} 
\usepackage{algorithm}
\usepackage{bbm}
\usepackage[table, dvipsnames]{xcolor}
\usepackage{enumitem}
\usepackage{rotating}
\usepackage{tikz}
\usetikzlibrary{shapes.geometric, arrows}
\usetikzlibrary{automata,positioning}
\usepackage{multirow}
\def\bSig\boldsymbol{\Sigma}

\usepackage{algorithm}
\usepackage{algpseudocode}

\usepackage{url}
\usepackage{cellspace}
\usepackage{subcaption}
\usepackage[font=bf]{caption}
\usepackage[colorlinks=true, linkcolor=black, urlcolor = black, citecolor=black]{hyperref} 
\usepackage[noabbrev,capitalize]{cleveref}

\providecommand{\keywords}[1]
{
  \small	
  \textbf{\textit{Keywords---}} #1
}
\newcommand{\be}{\begin{eqnarray}}
\newcommand{\ee}{\end{eqnarray}}
\newcommand{\bee}{\begin{eqnarray*}}
\newcommand{\eee}{\end{eqnarray*}}
\newcommand{\bi}{\begin{enumerate}}
\newcommand{\ei}{\end{enumerate}}

\crefformat{equation}{(#2#1#3)}

\title{Derivation of outcome-dependent dietary patterns for low-income women obtained from survey data using a Supervised Weighted Overfitted Latent Class Analysis}
\author{Stephanie M. Wu $^{1,*}$, 
Matthew R. Williams $^{2,**}$,  
Terrance D. Savitsky $^{3,***}$, \\
Briana J.K. Stephenson $^{1,****}$\\
\\
\footnotesize $^{1}$Department of Biostatistics, Harvard T.H. Chan School of Public Health, Boston, Massachusetts, U.S.A \\
\footnotesize $^{2}$RTI International, Research Triangle Park, North Carolina, U.S.A\\
\footnotesize $^{3}$Office of Survey Methods Research, U.S. Bureau of Labor Statistics, Washington, DC, U.S.A
\\
\footnotesize *\textit{email:} swu@g.harvard.edu \\
\footnotesize *\textit{email:} mrwilliams@rti.org \\
\footnotesize *\textit{email:} savitsky.terrance@bls.gov\\
\footnotesize *\textit{email:} bstephenson@hsph.harvard.edu
}
\date{} 

\begin{document}\emergencystretch 3em
\maketitle

\singlespacing
\begin{abstract}
Poor diet quality is a key modifiable risk factor for hypertension and disproportionately impacts low-income women. Analyzing diet-driven hypertensive outcomes in this demographic is challenging due to the complexity of dietary data and selection bias when the data come from surveys, a main data source for understanding diet-disease relationships in understudied populations. Supervised Bayesian model-based clustering methods summarize dietary data into latent patterns that holistically capture  relationships among foods and a known health outcome but do not sufficiently account for complex survey design. This leads to biased estimation and inference and lack of generalizability of the patterns. To address this, we propose a supervised weighted overfitted latent class analysis (SWOLCA) based on a Bayesian pseudo-likelihood approach that integrates sampling weights into an exposure-outcome model for discrete data. Our model adjusts for stratification, clustering, and informative sampling, and handles modifying effects via interaction terms within a Markov chain Monte Carlo Gibbs sampling algorithm. Simulation studies confirm that the SWOLCA model exhibits good performance in terms of bias, precision, and coverage. Using data from the National Health and Nutrition Examination Survey (2015-2018), we demonstrate the utility of our model by characterizing dietary patterns associated with hypertensive outcomes among low-income women in the United States. 
\end{abstract} \hspace{10pt}

%
\noindent

\clearpage
\section{Introduction}
\label{s:intro}
Low-income women are understudied in cardiometabolic research despite being disproportionately burdened by poor diet quality and its negative health impacts \citep{zhang2018trends}. Hypertension, a pervasive and major risk factor for cardiovascular disease, illustrates this gap \citep{whelton20182017}. While considerable research explores diet-hypertension links, few cohort studies focus on low-income women. Therefore, analyzing diet-hypertension patterns in this key demographic requires greater reliance on data from surveys, which allow targeted inclusion of hard-to-reach communities through techniques such as oversampling and stratification. When using survey data, analyses must properly account for complex survey design elements to avoid biased estimation and variance underestimation and to generalize beyond the sample \citep{pfeffermann1996use, parker2022computationally, williams2021uncertainty}. 

Dietary scores, such as the Dietary Approaches to Stop Hypertension (DASH) score \citep{sacks2001effects}, have been used to evaluate intake of key food groups according to prescriptive guidance. These metrics are standardized across populations but can lack flexibility in grouping foods in ways more reflective of population-specific dietary behavior. Alternatively, latent class analysis (LCA) is a clustering method that achieves this level of flexibility through data-driven derivation of underlying dietary consumption patterns \citep{lazarsfeld1968latent, sotres2010latent}. This enables additional insight into the behaviors of targeted populations and the creation of policy tailored to their dietary needs. Exploration of diet-disease relationships using LCA typically entails a two- or three-step approach. First, LCA is used to identify patterns; then, an association is measured via regression analysis using the LCA-derived pattern as a covariate, with possible bias adjustments to account for measurement error \citep{fung2001dietary, bray2015eliminating}. This is useful when testing a single exposure across many outcomes; however, when interest lies in obtaining a targeted understanding of how dietary patterns influence a specific health outcome, such as hypertension, a one-step supervised approach offers advantages in identifying outcome-informed patterns and smaller diet-outcome effects while correctly propagating classification uncertainty \citep{stephenson2022derivation, molitor2010bayesian, elliott2020methods}.

Extensions of LCA-based approaches to account for survey design have been met with challenges. Under a frequentist setting, high-dimensionality and sparseness of diet data lead to issues with parameter stability and matrix inversion  \citep{asparouhov2005sampling,patterson2002latent}. Under a Bayesian setting, models lack proper variance estimation \citep{stephenson2023racial, stephenson2024identifying}, inhibit classification \citep{gunawan2020bayesian}, or ignore the survey design entirely. Without proper incorporation of design features such as informative sampling and clustering, dietary patterns can be misidentified, and characterization of the diet-hypertension relationship can be biased with incorrect posterior intervals.

This paper aims to improve analysis of diet-hypertension patterns in low-income women using survey data. We propose a supervised weighted overfitted latent class analysis (SWOLCA) that uses a Bayesian pseudo-likelihood approach to account for complex survey design and produce accurate estimation and uncertainty quantification for a multivariate categorical exposure and a binary outcome. We also introduce a mixture reference coding scheme to allow interactions between dietary patterns and other covariates. Our model enables us to: 1) uncover the prevalence and profile of dietary patterns dependent on hypertensive status amongst low-income adult women; 2) efficiently measure the association between diet and hypertension while accounting for interactions with covariates; and 3) integrate survey sampling weights to produce accurate point and interval estimation for our target population.

The remaining sections of this paper are organized as follows. In Section~\ref{s:model}, we describe our proposed SWOLCA model along with a brief background. In Section~\ref{s:param}, we discuss implementation considerations for parameter estimation. In Section~\ref{s:sim}, we conduct a simulation study comparing SWOLCA with existing methods. In Section~\ref{s:app}, we apply the model to data from the National Health and Nutrition Examination Survey (NHANES) to describe dietary pattern association with hypertension among low-income women in the US. Finally, in Section~\ref{s:discuss}, we provide concluding remarks and discussion. 

\section{Model}
\label{s:model}

\subsection{Supervised Overfitted Latent Class Analysis}

Supervised overfitted latent class analysis (SOLCA) is a Bayesian nonparametric mixture model that jointly estimates latent dietary patterns through an overfitted latent class model and their associations to a binary hypertension outcome through a probit regression model. In this way, the latent patterns are informed by both the multivariate categorical diet exposure, $\boldsymbol{x}_{i\cdot}=(x_{i1}, \ldots, x_{iJ})^T$ for $J$ food items, and the binary hypertension outcome, $y_i$, which can increase precision and reduce bias. Each sampled individual $i \in \{1,\ldots, n\}$ is assigned to a dietary pattern $c_i \in \{1,\ldots, K\}$, where $K$ is the number of patterns. Model parameters include: the dietary pattern prevalences, $\boldsymbol{\pi} = (\pi_1, \ldots, \pi_K)^T$, where $\sum_{k=1}^K \pi_k =1$; the food item consumption probabilities characterizing the pattern compositions, $(\theta_{jc_i1}, \ldots, \theta_{jc_iR_j})$, where $\sum_{r=1}^{R_j}\theta_{jc_ir}=1$ for food item $j$ with $R_j$ consumption levels; and the probit regression coefficients, $\boldsymbol{\xi}_{c_i\cdot} = (\xi_{c_i1}, \ldots, \xi_{c_iq})^T$, corresponding to $q$ regression covariates, $\boldsymbol{v_i}$, given assignment to dietary pattern $c_i$, which is an unknown latent random variable simultaneously determined by the model. We use the formulation of the probit regression model introduced by \cite{albert1993bayesian}, where $z_i$ is a latent Gaussian variable such that $z_i \sim N(\boldsymbol{v}_i^T \boldsymbol{\xi}_{c_i\cdot}, 1)$, and is truncated depending on binary outcome $y_i$ so that $z_i>0$ when $y_i=1$ and $z_i\leq 0$ otherwise. Using this formulation, the SOLCA complete data joint distribution of ($\boldsymbol{x}_i, c_i, y_i, z_i$) is:
\begin{align}\label{e:solca_new}
p(\boldsymbol{x}_{i\cdot},c_i, y_i,z_i&|\boldsymbol{v}_i, \boldsymbol{\pi}, \boldsymbol{\theta}, \boldsymbol{\xi})\nonumber\\
&= p(c_i|\boldsymbol{\pi})p(\boldsymbol{x}_i|c_i, \boldsymbol{\theta})p(y_i, z_i|c_i,\boldsymbol{v}_i, \boldsymbol{\xi})\nonumber\\
&=\pi_{c_i}\prod_{j=1}^J\prod_{r=1}^{R_j}\theta_{jc_ir}^{I(x_{ij}=r)} \frac{e^{-\frac{1}{2}(z_i-\boldsymbol{v}_i^\intercal \boldsymbol{\xi}_{c_i\cdot})^2}}{\sqrt{2\pi}}\Big\{y_iI(z_i>0) + (1-y_i)I(z_i\leq 0)\Big\},
\end{align}
where $\boldsymbol{\xi} = (\boldsymbol{\xi}_{1\cdot}^T, \ldots, \boldsymbol{\xi}_{K\cdot}^T)^T$ is a $K \times q$ matrix of regression coefficients, $I(A)$ is the indicator function equal to 1 if $A$ is true and 0 otherwise, and $\boldsymbol{\theta}$ is a $J\times K \times R$ array with cells $\theta_{jkr}$, where $R=\max_jR_j$. SOLCA assumes items are independent conditional on dietary pattern assignment and individuals with the same pattern share behaviors for all food items.

The overfitted formulation of SOLCA enables a data-driven approach to select the number of patterns, $K$, without need for post-hoc testing \citep{van2015overfitting}. $K$ is set to a conservatively high number to allow empty patterns to drop out via a sparsity-inducing Dirichlet prior: $(\pi_1, \ldots, \pi_K) \sim \text{Dir}(\alpha_1, \ldots, \alpha_K)$, where hyperparameters $\alpha_k$ moderate the rate of growth for nonempty patterns and reduce the dependence of $K$ on the sample size and data structure. Smaller values of $\alpha_k$ yield a slower growth rate and more sparsity. 

\subsection{Supervised Weighted Overfitted Latent Class Analysis for Survey Data}

Supervised weighted overfitted latent class analysis (SWOLCA) is an extension of SOLCA to the survey setting. To obtain unbiased estimation of our target population, survey sampling weights for all sampled individuals are necessary. Stratification and clustering information of the survey are also needed for accurate variance estimation. We follow a weighted pseudo-likelihood approach as described in \cite{savitsky2016bayesian} and \cite{kunihama2016nonparametric}. Survey weights are used to up-weight individual likelihood contributions proportional to the number of individuals represented in the target population. This forms a weighted pseudo-likelihood that is used in place of the likelihood in the posterior update. Estimation and inference proceed using the posterior density of model parameters. Let $w_i$ denote the survey weight of individual $i$, $i\in\{1,\ldots, n\}$.  We use a normalization constant $\kappa = \sum_{i=1}^n w_i/n$ so the weights sum to $n$ to reflect sampling variability. This provides a coarse adjustment for the posterior uncertainty that can be further refined in a post-processing step described below. Denote all parameters and complete data of the unweighted SOLCA model with $\boldsymbol{\Theta}$ and $\boldsymbol{D}$, respectively. Then, the posterior density for the weighted SWOLCA approach is 
\begin{equation}\label{e:pseudo}
\widetilde{p}(\boldsymbol{\Theta} | \boldsymbol{D}) \propto p(\boldsymbol{\Theta})\prod_{i=1}^n p(\boldsymbol{D}_i|\boldsymbol{\Theta})^{\frac{w_i}{\kappa}}.
\end{equation}

Under certain regularity conditions, the posterior is consistent under unequal probability sampling \citep{savitsky2016bayesian} and complex multi-stage sampling \citep{williams2021uncertainty, williams2020bayesian}. However, posterior credible intervals will exhibit undercoverage due to clustering and population generation uncertainty that are not accounted for  \citep{leon2019fully, gunawan2020bayesian}. To address this, we extend the post-processing adjustment proposed in \cite{williams2021uncertainty} to accommodate a mixture model setting with constrained parameters. Posterior samples are rescaled to recover the correct ``sandwich" form of the asymptotic variance based on pseudo-MLE theory. Let $\widehat{\boldsymbol{\Theta}}_m$ denote the posterior estimates for Markov chain Monte Carlo (MCMC) sample $m$, with mean $\overline{\boldsymbol{\Theta}}$ across all samples. The rescaled estimates are 
\begin{equation}
    \widehat{\boldsymbol{\Theta}}_m^a = \Big(\widehat{\boldsymbol{\Theta}}_m - \overline{\boldsymbol{\Theta}}\Big) \boldsymbol{R_2}^{-1}\boldsymbol{R_1} + \overline{\boldsymbol{\Theta}},
\end{equation}
where $\boldsymbol{R_1}^T\boldsymbol{R_1}$ is the correct asymptotic ``sandwich" covariance of the pseudo-MLE and $\boldsymbol{R_2}^T \boldsymbol{R_2}$ is the asymptotic covariance of the posterior. $\boldsymbol{R_1}$ can be obtained using a mix of resampling and computing the posterior Hessian matrix. For the post-processing adjustment, using the normalization constant $\kappa$ for the sampling weights is not strictly necessary for correct uncertainty coverage but can improve numerical stability when computing $\boldsymbol{R_1}$. Some alternative variations of $\kappa$ may lead to smaller post-processing adjustments being needed, for example using an effective sample size based on variation of the weights \citep{spencer2000approximate}. 

\section{Parameter Estimation}
\label{s:param}
\subsection{MCMC Computation}

For SWOLCA parameter estimation, we implement a MCMC Gibbs sampling algorithm. We follow \cite{moran2021bayesian} and implement sampling in a two-stages: (1) an adaptive sampler estimates the appropriate number of dietary patterns, and (2) a fixed sampler generates model estimates based on the estimated number of patterns. Derivations of the full conditionals are provided in the Supplementary Materials. Proper mixing is encouraged via a random permutation sampler that is incorporated in the MCMC sampling algorithm \citep{fruhwirth2001markov}. Stan \citep{carpenter2017stan} is used in the calculation of the post-processing variance adjustment, as it offers automatic differentiation capabilities to compute the posterior gradient and Hessian. Due to issues with handling discrete latent variables in mixture model settings, Stan was not implemented for parameter sampling. 

\subsection{Mixture Reference Coding of Parameters}

A common concern with mixture models is label switching
\citep{stephens2000dealing}. Under a reference cell coding scheme, label switching turns the intercept and slope coefficients into noise due to switches in the reference pattern. Alternative coding schemes that have been used do not consider pattern-by-covariate interactions and lead to restrictions of the parameter space that are difficult to interpret under a probit link function \citep{molitor2010bayesian, stephenson2022derivation}. We resolve this by introducing a combination of factor variable \citep{buis2012stata} and reference cell coding, hereafter referred to as “mixture reference coding.” In mixture reference coding, the different dietary patterns are expressed in factor variable form, while the levels of any additional covariates are expressed in reference cell form. For example, suppose $c_i \in \{1, 2, 3\}$ is individual $i$'s dietary pattern assignment and $v_i \in \{0, 1\}$ is a binary covariate. Mixture reference coding for the probit regression model is given by:
\begin{align}\label{e:probit}
    \mathbb{E}(y_i|c_i, v_i) &= \Phi\Big\{\xi_{11}I(c_i=1) + \xi_{12}I(c_i=1)v_i \nonumber\\
    &\qquad + \xi_{21}I(c_i=2) + \xi_{22}I(c_i=2)v_i\nonumber\\
    &\qquad + \xi_{31}I(c_i=3) + \xi_{32}I(c_i=3)v_i\Big\}.
\end{align}

\noindent Essentially, each dietary pattern has its own reference parameter and corresponding regression model. All interactions between dietary pattern and additional covariates are captured, and additional interactions between covariates can be specified if desired. This balances flexibility, by allowing for interactions, with parsimony, by not forcing inclusion of all interactions between variables. It also allows any dietary pattern to be set as the reference level post-hoc. Using mixture reference coding, label switching can be resolved by adapting a post-processing hierarchical clustering relabeling approach \citep{krebs1999ecological, medvedovic2002bayesian, stephenson2022derivation}. 

\section{Simulation Study}
\label{s:sim}
\subsection{Simulation Design}
We conduct a simulation study to assess whether the proposed SWOLCA is able to produce valid estimation and inference of a target population sampled under complex survey designs. Our parameters of interest are the number of dietary patterns, $K$, their estimated prevalences, $\boldsymbol{\pi}$, the composition of each pattern,  $\boldsymbol{\theta}$, and the associations between the patterns and the observed outcome, $\boldsymbol{\xi}$. We compare our method to two alternatives: 1) an unweighted SOLCA that ignores survey design, and 2) a two-step approach where the first step fits an unsupervised weighted overfitted latent class analysis (WOLCA) to derive the dietary patterns \citep{stephenson2024identifying}, and the second step treats the pattern assignments as fixed and includes them as covariates in a survey-weighted regression model using R \texttt{survey} package version 4.1.1  \citep{lumley2004analysis}. All models are implemented in R version 4.2.0 \citep{RCoreTeam2023R} with C++ interface using the \texttt{Rcpp} package version 1.0.10 \citep{eddelbuettel2011journal}.  We run a Gibbs sampler for 20,000 iterations with 10,000 burn-in and thinning every 5 iterations. 

Data are generated for a finite population of size $N = 80,000$ with a total of $K=3$  dietary patterns that are also associated with a binary outcome.  Each pattern consists of $J=30$ categorical food items, consumed at one of $R=4$ levels. Survey features in the data include clustered outcomes and two unequal-sized strata in the population, with stratum membership influencing dietary pattern membership and the probability of the outcome. Full details of the data generation process are provided in the Supplementary Materials.

Model performance is evaluated for the sampling and data generating scenarios provided in Table~\ref{t:sims}. We examine three survey designs: simple random sampling (SRS); stratified sampling with unequal sampling probabilities; or stratified cluster sampling with unequal sampling probabilities and correlated outcomes. We focus on two associations of interest: a conditional outcome model with stratum included as a covariate; or a marginal outcome model that does not condition on selection or adjust for selection bias. And we compare three different sample sizes: 1\% (n = 800), 5\% (n = 4000), or 10\% of the population (n = 8000). Bold text indicates deviation from the default setting (scenario 2) of stratified sampling with a conditional model and sample size 4000. Model robustness is also evaluated in cases where a) additional confounders are included, b) latent patterns are defined with weak identifiability, and c) weakly separated patterns are defined with a few differing exposure variables driving the true association to the outcome. Descriptions and results for these additional scenarios are not shown here but are detailed in the Supplementary Materials.  

100 simulated datasets are generated for each scenario. Models are initialized with $K=30$ and Dirichlet hyperparameter  $\alpha = 1/K$ for all $k \in \{1\ldots, K\}$ to encourage sparsity and moderate growth of new pattern formation \citep{van2015overfitting}. A noninformative flat Dir(1) prior is used for $\boldsymbol{\theta}_{jc_i\cdot}$, and weakly informative priors are used for the regression parameters $\boldsymbol{\xi}$. To compare model performance for parameter estimation, we examine mean absolute bias (mean absolute distance between estimated and true parameter values), variability (full width of the 95\% credible interval (CI), averaged over dietary patterns), and coverage (proportion of 95\% CIs that cover the true population parameter values, averaged over dietary patterns).  

\subsection{Simulation Results}

\begin{table}
\small
\caption{\textmd{Absolute bias, 95\% credible interval width, and coverage for the unweighted SOLCA, two-step WOLCA, and proposed SWOLCA, based on posterior MCMC samples and averaged across 100 independent draws from the population. Strat = stratified sampling, Strat Cl = stratified cluster sampling, Cond = conditional model, Marg = marginal model. Notable issues of bias, imprecision, and undercoverage are underlined to improve readability.}}
\centering
\setlength{\tabcolsep}{0.33em}
\setlength\cellspacetoplimit{5pt}
\setlength\cellspacebottomlimit{5pt}
\begin{tabular}[t]{llcccccccccc}
\toprule
& & \multicolumn{4}{Sc}{\textbf{Absolute Bias}} & \multicolumn{3}{c}{\textbf{CI Width}} & \multicolumn{3}{c}{\textbf{Coverage}}\\
\cmidrule(lr){3-6}
\cmidrule(lr){7-9}
\cmidrule(l){10-12}
\textbf{Scenario} & \textbf{Model} & $\boldsymbol{K}$ & $\boldsymbol{\pi}$ & $\boldsymbol{\theta}$ & $\boldsymbol{\xi}$ & $\boldsymbol{\pi}$ & $\boldsymbol{\theta}$ & $\boldsymbol{\xi}$ & $\boldsymbol{\pi}$ & $\boldsymbol{\theta}$ & $\boldsymbol{\xi}$ \\
\midrule
(1) \textbf{SRS}, Cond, n=4000 & SOLCA & 0.00 & 0.006 & 0.006 & 0.063 & 0.027 & 0.042 & 0.367 & 0.957 & 0.958 & 0.965\\
& WOLCA & 0.00 & 0.006 & 0.006 & 0.063 & 0.036 & 0.044 & 0.762 & 0.950 & 0.958 & 0.992\\
& SWOLCA & 0.00 & 0.006 & 0.006 & 0.063 & 0.027 & 0.042 & 0.419 & 0.947 & 0.953 & 0.983\\[1ex]
(2) \textbf{Strat}, Cond, n=4000 & SOLCA & 0.00 & \underline{0.081} & 0.006 & 0.047 & \underline{0.069} & 0.045 & 0.374 & \underline{0.190} & 0.962 & 0.972\\
& WOLCA & 0.00 & 0.006 & 0.007 & 0.043 & 0.031 & 0.045 & 0.672 & 0.957 & 0.933 & 0.998\\
& SWOLCA & 0.00 & 0.006 & 0.006 & 0.044 & 0.036 & 0.049 & 0.414 & 0.977 & 0.952 & 0.990\\[1ex]
(3) \textbf{Strat Cl}, Cond, n=4000 & SOLCA & 0.00 & \underline{0.082} & 0.006 & 0.132 & \underline{0.074} & 0.046 & 0.390 & \underline{0.223} & 0.966 & \underline{0.592}\\
& WOLCA & 0.00 & 0.006 & 0.006 & 0.127 & 0.037 & 0.044 & \underline{1.210} & 0.963 & 0.942 & 0.990\\
& SWOLCA & 0.00 & 0.006 & 0.006 & 0.126 & 0.031 & 0.047 & 0.816 & 0.950 & 0.942 & 0.963\\[1ex]
(4) Strat, \textbf{Marg}, n=4000 & SOLCA & 0.00 & 0.008 & 0.006 & \underline{0.203} & 0.062 & 0.043 & 0.162 & 0.963 & 0.958 & \underline{0.063}\\
& WOLCA & 0.00 & 0.016 & 0.007 & 0.031 & 0.107 & 0.049 & 0.348 & 0.947 & 0.939 & 0.993\\
& SWOLCA & 0.00 & 0.011 & 0.007 & 0.033 & 0.097 & 0.063 & 0.278 & 0.967 & 0.965 & 0.987\\[1ex]
(5) Strat, Cond, \textbf{n=8000} & SOLCA & 0.00 & \underline{0.080} & 0.005 & 0.049 & \underline{0.076} & 0.042 & 0.367 & \underline{0.227} & 0.972 & 0.980\\
& WOLCA & \underline{0.06} & 0.010 & 0.011 & 0.038 & 0.044 & 0.044 & 0.519 & 0.920 & \underline{0.908} & 0.960\\
& SWOLCA & 0.00 & 0.004 & 0.005 & 0.030 & 0.029 & 0.038 & 0.373 & 0.967 & 0.953 & 0.997\\[1ex]
(6) Strat, Cond, \textbf{n=800} & SOLCA & 0.00 & \underline{0.084} & 0.013 & 0.098 & 0.064 & 0.088 & 0.701 & \underline{0.027} & 0.938 & 0.945\\
& WOLCA & 0.00 & 0.013 & 0.014 & 0.099 & 0.060 & 0.095 & \underline{1.371} & 0.933 & 0.919 & 0.983\\
& SWOLCA & 0.00 & 0.013 & 0.014 & 0.097 & 0.062 & 0.099 & 0.687 & 0.947 & 0.922 & 0.947\\
\bottomrule
\end{tabular}
\label{t:sims}
\end{table}

For all models and scenarios, investigation of traceplots and autocorrelation plots showed good mixing and convergence of all model parameters. Table \ref{t:sims} displays a summary of simulation results for the scenarios described. As expected, under the control SRS scenario, all three models exhibit good estimation and coverage properties. For other scenarios with a variety of complex survey design and data-generating features, the proposed SWOLCA outperforms the two alternative models and is able to obtain accurate and precise estimation, as well as approximately nominal coverage, for all parameters.  

The unweighted SOLCA model gives highly biased estimates of the pattern membership probabilities $\boldsymbol{\pi}$ when there is stratified sampling. When there is cluster sampling, credible intervals for the regression coefficients $\boldsymbol{\xi}$ exhibit severe undercoverage, which can result in overconfident estimation of associational effects. When the selection is associated with hypertension and a marginal model is fit, SOLCA yields biased estimates for $\boldsymbol{\xi}$. 

The two-step WOLCA model produces estimates of regression parameters $\boldsymbol{\xi}$ that have wide credible intervals and are less precise than the SWOLCA model at similar coverage levels. This inefficiency is especially true for small sample sizes and cluster sampling designs because the two-step process ignores uncertainty in the first step. This inflates interval widths and makes inference on the true associational effects difficult. WOLCA is also the most prone to undercoverage of $\theta$ due to failure to account for variability in the plug-in survey weights in the first step, and it runs into issues with estimating the number of dietary patterns, $K$.

SWOLCA yields estimates with minimal bias and approximately nominal interval coverage for all parameters for stratified and cluster sampling designs. It is also able to use the survey weights to account for bias from selection variables that are unavailable for analysis, enabling correct marginal estimation of $\xi$ and producing outcome probability estimates that accommodate informative designs without greatly inflating uncertainty (Web Figure 2). In the cluster sampling and 1\% sample size scenarios, there is slight undercoverage of $\boldsymbol{\theta}$. This is expected given the increased variability of the data and is also seen in the SOLCA and WOLCA comparison models. These conclusions were consistent in settings with weaker patterns (mode 55\%), overlapping patterns where consumption of many foods is the same for two patterns, different sample sizes, and additional regression covariates. 

\section{Application to NHANES Low-Income Women}
\label{s:app}

\subsection{Data Description and Model Setup}
The National Health and Nutrition Examination Survey (NHANES) is a cross-sectional, nationally-representative survey that assesses the health and nutritional status of the non-institutionalized civilian US population \citep{nchs2023homepage}. The survey employs a stratified, clustered, four-stage sampling design with oversampling to increase inclusion of various age, sex, income, and racial and ethnic groups. Data are publicly available alongside survey sampling weights that take into account unequal sampling probabilities, stratification, clustering, non-response, weight trimming, and calibration \citep{chen2020national}. Data are pooled from two survey cycles, 2015-2016 and 2017-2018, in accordance with protocols outlined in the NHANES analytic guidelines \citep{national2018national}. We focus on dietary patterns associated with hypertension among adult females aged 20 or over who are classified as low-income (reported household income at or below 185\% of the federal poverty level, consistent with eligibility requirements for federal assistance program participation \citep{oliveira2009wic}). Pregnant or breastfeeding women are excluded ($n=179$), resulting in a total sample size of $n=2003$. 

Dietary exposure variables are defined as 28 food item groups collected from two 24-hour dietary recalls and summarized into food pattern equivalents from the Food and Nutrition Database for Dietary Studies  \citep{dietary2015dietary, bowman2020food}. Each food item is categorized as none, low, medium, or high, based on relative tertiles of positive consumption \citep{sotres2013maternal, stephenson2023racial}. The binary observed outcome, hypertension, is defined as a composite measure of having an elevated blood pressure (BP) reading (systolic BP $> 130$ or diastolic BP $> 80$), self-reported  diagnosis, or use of hypertension-controlling medication. Age, race and ethnicity, current smoking status, and physical activity are included as potential confounders in our hypertensive outcome regression model. Web Table 3 displays summaries of these demographic characteristics by hypertension in the sample. 

We compare the proposed SWOLCA model and the unweighted SOLCA model in assessing diet-driven hypertension using survey data. Both models are initialized with the same priors used in the simulation study. Full details of posterior computation are provided in the Supplementary Materials. Estimation is obtained by fitting a Gibbs sampler of 20,000 iterations with 10,000 burn-in and thinning every 5 iterations, then summarized using posterior median estimates and 95\% credible intervals. Both adaptive and fixed sampler are run for 20,0000 iterations each using an Apple M1 Pro computer with 8 cores, with a computation time of roughly 75 minutes. If $\widehat{K}$ is set a priori and only the fixed sampler is run, the computation time is approximately 15 minutes, assuming the same computing power.

\subsection{Dietary Pattern Results}
\begin{figure}
    \centering
    \begin{subfigure}[t]{0.49\textwidth}
        \centering
        \includegraphics[width=\textwidth]{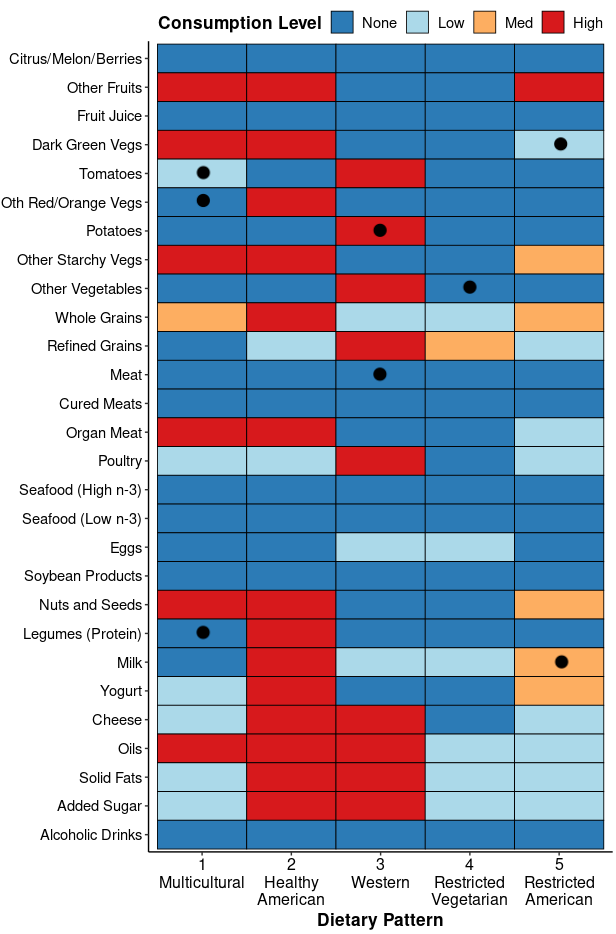}
        \caption{\textmd{SWOLCA}}
    \end{subfigure}
    \hfill
    \begin{subfigure}[t]{0.49\textwidth}
        \centering
        \includegraphics[width=\textwidth]{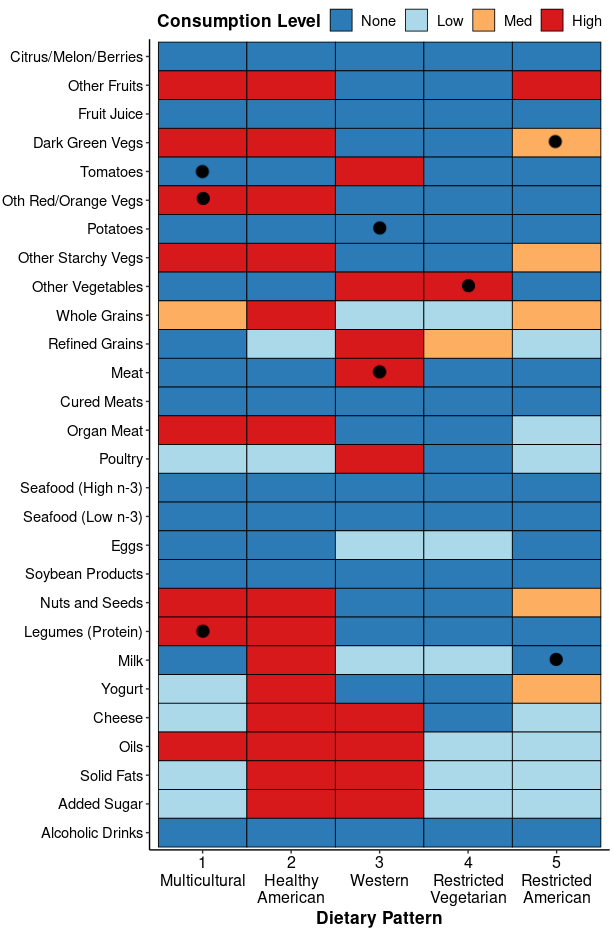}
        \caption{\textmd{SOLCA}}
    \end{subfigure}
    \caption{\textmd{Diet-hypertension patterns identified by the weighted SWOLCA and unweighted SOLCA models among low-income women in the US. Differences in modal consumption are indicated with black dots. Consumption levels are categorized as none, low, medium, and high. For each pattern, consumption of each food component is colored according to the modal consumption level (i.e., $\text{argmax}_r \theta_{jkr}$ for $r=1,\ldots, 4$, $j = 1,\ldots, 28$, $k = 1,\ldots, 5$).}}
    \label{f:theta_comparison}
\end{figure}

Both SWOLCA and SOLCA identify $\widehat{K}=5$ diet-hypertension patterns among low-income women in the US, displayed in Figure~\ref{f:theta_comparison} with characterization differences indicated by black dots. For all patterns, consumption behavior is colored by the modal (i.e., highest posterior probability) consumption level (none, low, medium, or high) for each of the 28 food items. We can see differences in modal food consumption for all patterns except Pattern 2 (Healthy American), illustrating the influence of survey weights on pattern composition. 

\begin{figure}
    \centering
    \includegraphics[width=\textwidth]{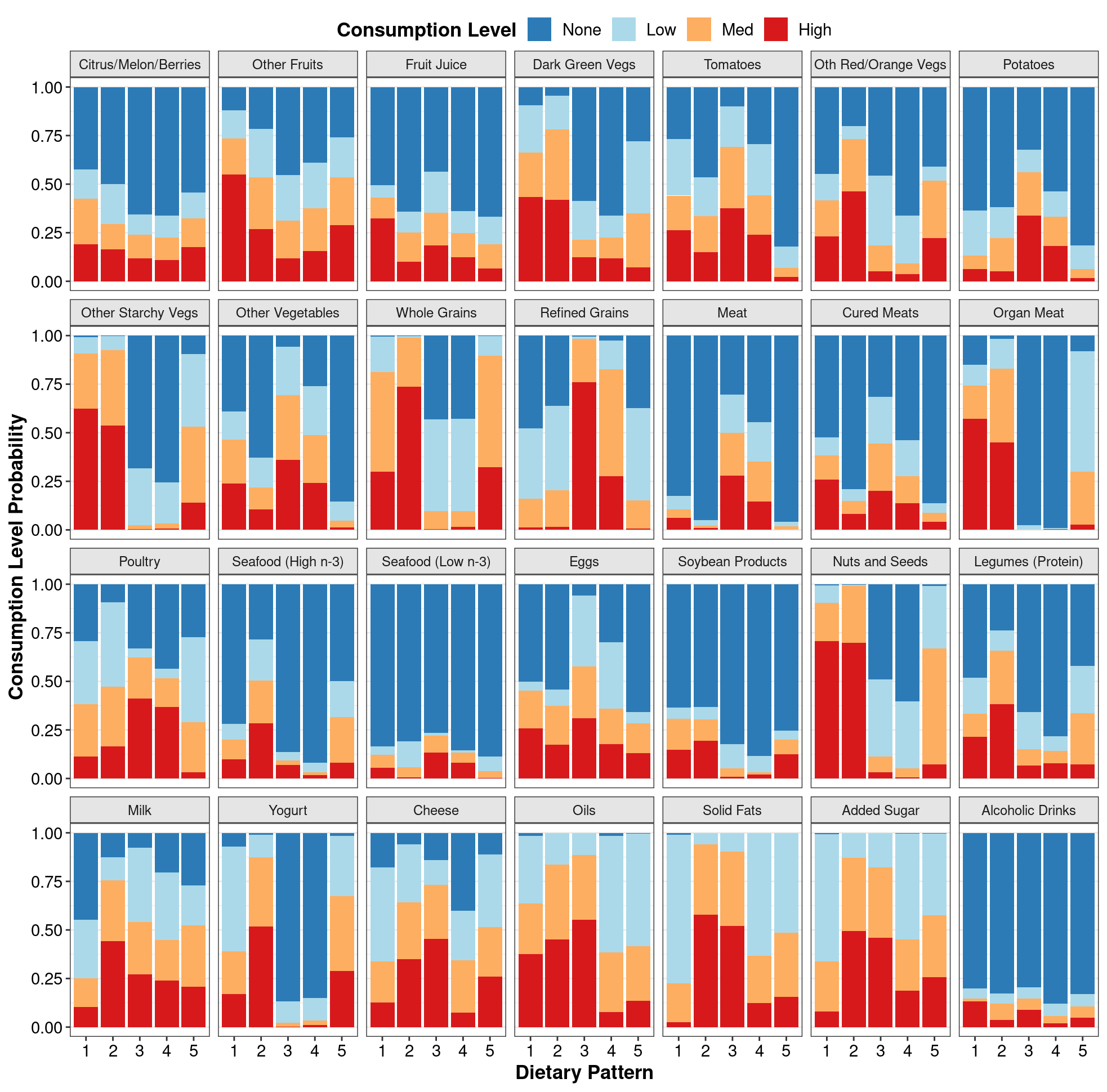}
    \caption{\textmd{Detailed breakdown of consumption level probabilities by diet-hypertension pattern for each food component for the five diet-hypertension patterns identified by the SWOLCA model among low-income women in the US. Pattern names: 1) Multicultural, 2) Healthy American, 3) Western, 4) Restricted Vegetarian, and 5) Restricted American.}}
    \label{f:theta_probs}
\end{figure}

We continue by focusing on the patterns identified by SWOLCA (Figure~\ref{f:theta_comparison}(a)). We characterize the five diet-hypertension patterns as follows: 1) Multicultural, 2) Healthy American, 3) Western, 4) Restricted Vegetarian, and 5) Restricted American. The Multicultural pattern favors high consumption of organ meat, starchy and dark green vegetables, oils, and fruits. It is referred to as the Multicultural pattern given its large prevalence among those identifying as NH Asian and groups other than NH White (Table~\ref{t:lc_demog}). The Healthy American pattern favors a higher consumption of healthier foods such as fruits, vegetables, whole grains, organ meat, and nuts, but still includes high consumption of oils, fats, and sugars prevalent in many American diets. The Western pattern favors a high consumption of refined grains, poultry, cheese, oils, solid fats, added sugars, and other vegetables. Examining Figure~\ref{f:theta_probs}, which provides detailed consumption level probabilities by pattern for each food item, we see that individuals assigned to the Western pattern are also the most likely to consume meat and cured meats, though this consumption is heterogeneous. The Restricted Vegetarian pattern favors no consumption of many foods including meat and seafood. There is moderate consumption of refined grains and low consumption of whole grains, poultry, eggs, milk, oils, solid fats, and added sugar. This population may face significant food access issues such as residence in or near a food desert or food swamp. Finally, the Restricted American pattern favors low consumption of many foods but to a lesser extent than the Restricted Vegetarian diet and with some intake of organ meats and poultry. This diet is similar to the Healthy American diet but with relatively lower consumption, especially for legumes and vegetables.

Examining the size and distribution of the diet-hypertension patterns across demographic variables (Table~\ref{t:lc_demog}), we see that the Western diet is most prevalent (26.7\% of the population) and the Multicultural diet is least prevalent (10.7\%). Those who follow the Multicultural diet tend to be younger, NH Asian, not a current smoker, and physically active. Conversely, those who follow the Restricted American diet tend to be older, NH White, and a current smoker. Those who follow the Restricted Vegetarian diet also tend to be older, current smokers, and physically inactive. 

\begin{table}
    \footnotesize
    \centering
    \caption{\textmd{Size and demographic distribution of dietary patterns for low-income women in the US. Estimates for $N$ use posterior samples of  parameter $\pi$. Column-wise mean and percentage estimates of demographic variables are calculated using sampling weights.}}
    \begin{tabular}{llllllll}
\toprule
Variable & Level & Multicultural & Healthy & Western & Restrict Veg & Restrict & Overall\\
\midrule
\multicolumn{2}{l}{$N$: \% (posterior SD \%)}  & 10.7 (4.2) & 24.4 (3.8) & 26.7 (3.9) & 19.9 (4.3) & 18.3 (5.4) & \\
\multicolumn{2}{l}{$n$: \%} & 11.5 & 23.6 & 25.8 & 20.7 & 18.4 & \\
\addlinespace
Age Group: \% & {}[20,40) & 51.7 & 47.1 & 44.8 & 33.2 & 35.3& 41.9\\
 & {}[40,60) & 28.6 & 29.1 & 33.4 & 37.1 & 33.1 & 32.6\\
 & $\geq 60$ & 19.6 & 23.9 & 21.9  & 29.7 & 31.6 & 25.5 \\
 \addlinespace
Race and Ethnicity: \% & NH White & 36.9 & 50.1 & 49.0 & 49.7 & 55.0 & 49.3\\
 & NH Black & 18.2 & 14.7 & 17.7 & 15.6 & 16.3 & 16.3\\
 & NH Asian & 20.2 & 2.2 & 3.4 & 7.1 & 3.4 & 5.6\\
 & Hispanic/Latino & 23.6 & 27.2 & 23.8 & 20.4 & 22.2 & 23.6\\
 & Other/Mixed & 1.2 & 5.8 & 6.1 & 7.2 & 3.1 & 5.2\\
 \addlinespace
Smoking Status: \% & Non-Smoker & 85.3 & 72.7 & 75.3 & 69.9 & 69.3 & 73.5\\
 & Smoker & 14.7 & 27.3 & 24.7 & 30.1 & 30.7 & 26.5\\
 \addlinespace
Physical Activity: \% & Inactive & 41.0 & 47.0 & 45.5 & 47.3 & 41.7 & 45.1\\
 & Active & 59.0 & 53.0 & 54.5 & 52.7 & 58.3 & 54.9\\
\bottomrule
\end{tabular}
    \label{t:lc_demog}
\end{table}

\subsection{Dietary Patterns and Hypertension Risk} 

Table~\ref{t:coefs_comparison} displays the posterior $\boldsymbol{\xi}$ estimates for the main dietary pattern effects on hypertension for SWOLCA and SOLCA. The full tables of regression estimates are provided in the Supplementary Materials. Estimates for the unweighted SOLCA differ greatly from those of SWOLCA, indicating presence of selection variables that influence the outcome but were not included as covariates. This results in inaccurate contributions of individuals to the estimation of hypertension probabilities because sampling weights have not been considered in the regression estimation process. The unweighted model produces fewer interaction effects and reduced ability to distinguish effects between the patterns. Credible intervals are also much tighter when the survey design is not incorporated. This leads to incorrect inference, as illustrated by the undercoverage shown in the simulation study.

\begin{table}
    \small
    \centering
    \caption{\textmd{Main effect probit regression parameter estimates for the proposed SWOLCA and the unweighted SOLCA, adjusting for demographic confounders. Reference group: Multicultural diet, age [20,40), NH White, non-smoker, inactive.}}
\begin{tabular}{l>{}l>{}l>{}l>{}l>{}l>{}l}
\toprule
& \multicolumn{3}{Sc}{\textbf{SWOLCA}} & \multicolumn{3}{c}{\textbf{SOLCA}}\\
\cmidrule(lr){2-4}
\cmidrule(lr){5-7}
Covariate & Estimate & 95\% CI & P($\xi > 0$ )& Estimate & 95\% CI  & P($\xi > 0$ )\\
\midrule
(Intercept)  & -1.66 &(-2.92, -0.41) & $<$0.01  & -1.01 &(-1.61, -0.41) & $<$0.01\\
Multicultural & 1 (reference) & -& -& -&- &-\\
Healthy Amer & 0.45 &(-1.12,  1.96) & 0.70 & 0.05 &(-0.68,  0.81) & 0.56\\
Western & 0.49 &(-0.86,  1.83) & 0.75 & 0.02 &(-0.71,  0.70) & 0.53\\
Restricted Veg & 0.67 &(-0.88,  2.17) & 0.79 & -0.11 &(-0.89,  0.64) & 0.39\\
Restricted Amer & 1.01 &(-0.21,  2.15) & 0.96 & 0.14 &(-0.65,  0.91) & 0.63\\
\bottomrule
\end{tabular}
    \label{t:coefs_comparison}
\end{table}

\begin{figure}
    \centering
    \includegraphics[width = \linewidth]{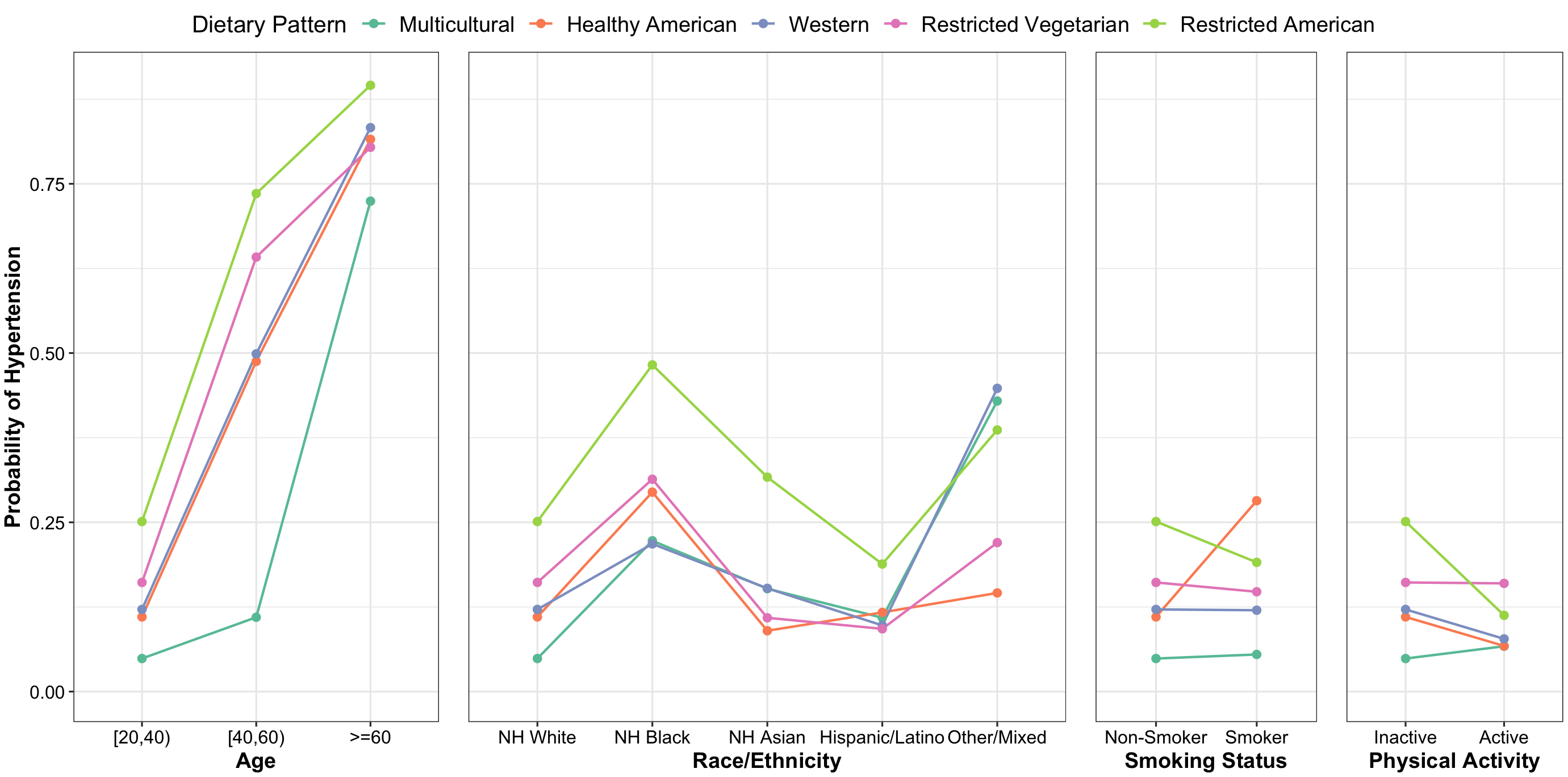}
    \caption{\textmd{Estimated probability of hypertension outcome by diet-hypertension pattern for all covariates, including interactions with pattern, for the SWOLCA model. For each covariate plot, all other covariates are set to the following  baseline values: Multicultural diet, age [20,40), NH White, non-smoker, and inactive.}}
    \label{f:phi}
\end{figure}

Focusing on the SWOLCA results, all diets appear to be associated with increased probability of hypertension compared to the Multicultural diet, with the Restricted diets showing the strongest increase (Table~\ref{t:coefs_comparison}). Figure~\ref{f:phi} displays hypertension probabilities of the diet-hypertension patterns as well as interactions effects between the patterns and socio-demographic variables. The estimated hypertension probabilities for the patterns were: 4.8\% (Multicultural), 11.1\% (Healthy American), 12.2\% (Western), 15.9\% (Restricted Vegetarian), and 25.2\% (Restricted American), among those aged 20 to 39, identifying as NH White, not currently smoking, and physically inactive. Age has a strong, positive association with elevated probability of hypertension, and the differentials among patterns is most pronounced in the 40 to 60 age group, with the Multicultural diet outcome probability remaining low (13\%) in the 40 to 60 age group, whereas the Restricted American diet sees a large increase in the 40 to 60 age group (74\%). Among different racial and ethnic groups, probability of hypertension is higher among those identifying as NH Black compared to NH White for all patterns, and the Restricted American pattern remains at high risk across all groups. The Healthy American pattern sees the largest increase in outcome probability among smokers, and the Restricted American pattern sees the largest decrease among those who are active.

\section{Discussion}
\label{s:discuss}

In this work, we develop the supervised weighted overfitted latent class analysis (SWOLCA), which is a Bayesian joint mixture model that can be used to: 1) elicit dietary patterns informed by both a set of categorical dietary intake exposures and a binary hypertension outcome, with a data-driven approach to determine the number of patterns; 2) capture small diet-hypertension effects and interactions with covariates via mixture reference coding; and 3) obtain unbiased estimation and inference for the population by adjusting for complex survey design features such as stratification, clustering, and informative sampling. Although our method is designed for categorical exposures and a binary outcome, extensions to other outcome data types can be made by replacing the probit likelihood with a regression likelihood that can accommodate different outcome data types (e.g., multinomial, ordinal, continuous). This remains an area of active research.  

Simulation studies confirmed that SWOLCA improved accuracy, precision, and coverage of parameter estimation compared to models that did not include sampling weights or relied on the two-step approach. Incorporation of survey design features was important for accurate estimation of the pattern prevalence, pattern identification, exposure-outcome association, and variance. Incorporation of the outcome into clustering improved precision of regression estimates and led to better identification of patterns. Implementation of SWOLCA to NHANES 2015-2018 data identified five diet-hypertension patterns among low-income US women. Differences between the SWOLCA and SOLCA results illustrated the importance of accounting for survey design. Failure to include survey sampling weights changed the pattern compositions because individual contributions to the consumption level probabilities during pattern formation reflected the sample composition and were not necessarily representative of the population. Our model identified strong age effects and captured substantial heterogeneity among different racial and ethnic subgroups via interaction terms.

Our work suggests several areas for further improvement. Firstly, our model does not adjust for data reliability typically encountered in dietary data, such as measurement error, recall bias, and item non-response missingness. Secondly, diet consumption heterogeneity may be better captured by adapting methods that allow demographic or behavior driven deviations of foods from the overall diet-disease pattern, and additionally incorporating survey design elements. Thirdly, our model is based on cross-sectional data and is not able to evaluate the impact of exposure changes over time or a time to event analysis. Lastly, our model relies on a probit regression component that can be limited by computational stability. Computation may be improved by incorporating hierarchical priors or by exploring other distributions, such as a unified skew normal conjugate model \citep{anceschi2023bayesian}. We leave these opportunities for extensions to future research.





\section*{Acknowledgements}

The authors are grateful to Walter Willett for helpful comments on earlier versions of this work, and to the Co-Editor, Associate Editor, and two referees for insightful comments that greatly improved the paper. This research was supported by the National Institute of Allergy and Infectious Diseases (NIAID: T32 AI007358) and the National Heart, Lung, and Blood Institute (NHLBI: R25 HL105400 awarded to Victor G. Davila-Roman and DC Rao). \vspace*{-8pt}

\section*{Supplementary Materials}

Web Appendices, Tables, and Figures, and data and code referenced in Sections \ref{s:model}, \ref{s:sim}, and \ref{s:app} are available with this paper at the Biometrics website on Oxford Academic. Code for replicating the simulations and data analyses in this paper is also available on GitHub at \url{https://github.com/smwu/SWOLCA} and is currently being developed into an R package. \vspace*{-8pt}

\section*{Data Availability Statement}
Data used in this paper to illustrate our findings are publicly available at \url{https://github.com/smwu/SWOLCA/} and were derived from the following resources available in the public domain: \url{https://wwwn.cdc.gov/nchs/nhanes/}.



\bibliographystyle{biom} \bibliography{aim1_biometrics}

\begin{thebibliography}{}

\bibitem[\protect\citeauthoryear{Albert and Chib}{Albert and
  Chib}{1993}]{albert1993bayesian}
Albert, J.~H. and Chib, S. (1993).
\newblock Bayesian analysis of binary and polychotomous response data.
\newblock {\em Journal of the American Statistical Association} {\bf 88,}
  669--679.

\bibitem[\protect\citeauthoryear{Anceschi, Fasano, Durante, and
  Zanella}{Anceschi et~al.}{2023}]{anceschi2023bayesian}
Anceschi, N., Fasano, A., Durante, D., and Zanella, G. (2023).
\newblock Bayesian conjugacy in probit, tobit, multinomial probit and
  extensions: A review and new results.
\newblock {\em Journal of the American Statistical Association} {\bf 118,}
  1451--1469.

\bibitem[\protect\citeauthoryear{Asparouhov}{Asparouhov}{2005}]{asparouhov2005sampling}
Asparouhov, T. (2005).
\newblock Sampling weights in latent variable modeling.
\newblock {\em Structural Equation Modeling} {\bf 12,} 411--434.

\bibitem[\protect\citeauthoryear{Bowman, Clemens, Friday, and Moshfegh}{Bowman
  et~al.}{2020}]{bowman2020food}
Bowman, S., Clemens, J., Friday, J., and Moshfegh, A. (2020).
\newblock Food patterns equivalents database 2017--2018: methodology and user
  guide.
\newblock {\em Food Surveys Research Group: Beltsville, MD} .

\bibitem[\protect\citeauthoryear{Bray, Lanza, and Tan}{Bray
  et~al.}{2015}]{bray2015eliminating}
Bray, B.~C., Lanza, S.~T., and Tan, X. (2015).
\newblock Eliminating bias in classify-analyze approaches for latent class
  analysis.
\newblock {\em Structural Equation Modeling} {\bf 22,} 1--11.

\bibitem[\protect\citeauthoryear{Buis}{Buis}{2012}]{buis2012stata}
Buis, M.~L. (2012).
\newblock Stata tip 106: With or without reference.
\newblock {\em The Stata Journal} {\bf 12,} 162--164.

\bibitem[\protect\citeauthoryear{Carpenter, Gelman, Hoffman, Lee, Goodrich,
  Betancourt, et~al\mbox{.}}{Carpenter et~al.}{2017}]{carpenter2017stan}
Carpenter, B., Gelman, A., Hoffman, M.~D., Lee, D., Goodrich, B., Betancourt,
  M., et~al. (2017).
\newblock Stan: A probabilistic programming language.
\newblock {\em Journal of Statistical Software} {\bf 76,} 1.

\bibitem[\protect\citeauthoryear{Chen, Clark, Riddles, Mohadjer, and
  Fakhouri}{Chen et~al.}{2020}]{chen2020national}
Chen, T.~C., Clark, J., Riddles, M.~K., Mohadjer, L.~K., and Fakhouri, T.~H.
  (2020).
\newblock National health and nutrition examination survey, 2015-2018: Sample
  design and estimation procedures.
\newblock {\em Vital and Health Statistics. Series 2, Data Evaluation and
  Methods Research} pages 1--35.

\bibitem[\protect\citeauthoryear{{Dietary Guidelines Advisory
  Committee}}{{Dietary Guidelines Advisory
  Committee}}{2015}]{dietary2015dietary}
{Dietary Guidelines Advisory Committee} (2015).
\newblock {\em Dietary Guidelines for Americans 2015-2020}.
\newblock Government Printing Office.

\bibitem[\protect\citeauthoryear{Eddelbuettel and Fran\c{c}ois}{Eddelbuettel
  and Fran\c{c}ois}{2011}]{eddelbuettel2011journal}
Eddelbuettel, D. and Fran\c{c}ois, R. (2011).
\newblock {Rcpp}: Seamless {R} and {C++} integration.
\newblock {\em Journal of Statistical Software} {\bf 40,} 1--18.

\bibitem[\protect\citeauthoryear{Elliott, Zhao, Mukherjee, Kanaya, and
  Needham}{Elliott et~al.}{2020}]{elliott2020methods}
Elliott, M.~R., Zhao, Z., Mukherjee, B., Kanaya, A., and Needham, B.~L. (2020).
\newblock Methods to account for uncertainty in latent class assignments when
  using latent classes as predictors in regression models, with application to
  acculturation strategy measures.
\newblock {\em Epidemiology} {\bf 31,} 194--204.

\bibitem[\protect\citeauthoryear{Fr{\"u}hwirth-Schnatter}{Fr{\"u}hwirth-Schnatter}{2001}]{fruhwirth2001markov}
Fr{\"u}hwirth-Schnatter, S. (2001).
\newblock Markov chain monte carlo estimation of classical and dynamic
  switching and mixture models.
\newblock {\em Journal of the American Statistical Association} {\bf 96,}
  194--209.

\bibitem[\protect\citeauthoryear{Fung, Willett, Stampfer, Manson, and Hu}{Fung
  et~al.}{2001}]{fung2001dietary}
Fung, T.~T., Willett, W.~C., Stampfer, M.~J., Manson, J.~E., and Hu, F.~B.
  (2001).
\newblock Dietary patterns and the risk of coronary heart disease in women.
\newblock {\em Archives of Internal Medicine} {\bf 161,} 1857--1862.

\bibitem[\protect\citeauthoryear{Gunawan, Panagiotelis, Griffiths, and
  Chotikapanich}{Gunawan et~al.}{2020}]{gunawan2020bayesian}
Gunawan, D., Panagiotelis, A., Griffiths, W., and Chotikapanich, D. (2020).
\newblock Bayesian weighted inference from surveys.
\newblock {\em Australian \& New Zealand Journal of Statistics} {\bf 62,}
  71--94.

\bibitem[\protect\citeauthoryear{Krebs}{Krebs}{1999}]{krebs1999ecological}
Krebs, C.~J. (1999).
\newblock {\em Ecological Methodology}.
\newblock Benjamin/Cummings, 2nd edition.

\bibitem[\protect\citeauthoryear{Kunihama, Herring, Halpern, and
  Dunson}{Kunihama et~al.}{2016}]{kunihama2016nonparametric}
Kunihama, T., Herring, A., Halpern, C., and Dunson, D. (2016).
\newblock Nonparametric bayes modeling with sample survey weights.
\newblock {\em Statistics \& Probability Letters} {\bf 113,} 41--48.

\bibitem[\protect\citeauthoryear{Lazarsfeld and Henry}{Lazarsfeld and
  Henry}{1968}]{lazarsfeld1968latent}
Lazarsfeld, P.~F. and Henry, N. (1968).
\newblock {\em Latent Structure Analysis}.
\newblock Houghton, Mifflin.

\bibitem[\protect\citeauthoryear{Le{\'o}n-Novelo and Savitsky}{Le{\'o}n-Novelo
  and Savitsky}{2019}]{leon2019fully}
Le{\'o}n-Novelo, L.~G. and Savitsky, T.~D. (2019).
\newblock Fully bayesian estimation under informative sampling.
\newblock {\em Electronic Journal of Statistics} {\bf 13,} 1608--1645.

\bibitem[\protect\citeauthoryear{Lumley}{Lumley}{2004}]{lumley2004analysis}
Lumley, T. (2004).
\newblock Analysis of complex survey samples.
\newblock {\em Journal of Statistical Software} {\bf 9,} 1--19.
\newblock R package version 2.2.

\bibitem[\protect\citeauthoryear{Medvedovic and Sivaganesan}{Medvedovic and
  Sivaganesan}{2002}]{medvedovic2002bayesian}
Medvedovic, M. and Sivaganesan, S. (2002).
\newblock Bayesian infinite mixture model based clustering of gene expression
  profiles.
\newblock {\em Bioinformatics} {\bf 18,} 1194--1206.

\bibitem[\protect\citeauthoryear{Molitor, Papathomas, Jerrett, and
  Richardson}{Molitor et~al.}{2010}]{molitor2010bayesian}
Molitor, J., Papathomas, M., Jerrett, M., and Richardson, S. (2010).
\newblock Bayesian profile regression with an application to the national
  survey of children's health.
\newblock {\em Biostatistics} {\bf 11,} 484--498.

\bibitem[\protect\citeauthoryear{Moran, Dunson, Wheeler, and Herring}{Moran
  et~al.}{2021}]{moran2021bayesian}
Moran, K.~R., Dunson, D., Wheeler, M.~W., and Herring, A.~H. (2021).
\newblock Bayesian joint modeling of chemical structure and dose response
  curves.
\newblock {\em The Annals of Applied Statistics} {\bf 15,} 1405--1430.

\bibitem[\protect\citeauthoryear{{National Center for Health
  Statistics}}{{National Center for Health
  Statistics}}{2018}]{national2018national}
{National Center for Health Statistics} (2018).
\newblock National health and nutrition examination survey: Analytic
  guidelines, 2011--2014 and 2015--2016.
\newblock Technical report, Centers for Disease Control and Prevention.

\bibitem[\protect\citeauthoryear{{National Center for Health
  Statistics}}{{National Center for Health
  Statistics}}{2023}]{nchs2023homepage}
{National Center for Health Statistics} (2023).
\newblock National health and nutrition examination survey home page.

\bibitem[\protect\citeauthoryear{Oliveira and Fraz{\~a}o}{Oliveira and
  Fraz{\~a}o}{2015}]{oliveira2009wic}
Oliveira, V. and Fraz{\~a}o, E. (2015).
\newblock The wic program: Background, trends, and economic issues, 2015
  edition. economic information bulletin number 134.
\newblock Technical report, US Department of Agriculture, Economic Research
  Service.

\bibitem[\protect\citeauthoryear{Parker, Holan, and Janicki}{Parker
  et~al.}{2022}]{parker2022computationally}
Parker, P.~A., Holan, S.~H., and Janicki, R. (2022).
\newblock Computationally efficient bayesian unit-level models for non-gaussian
  data under informative sampling with application to estimation of health
  insurance coverage.
\newblock {\em The Annals of Applied Statistics} {\bf 16,} 887--904.

\bibitem[\protect\citeauthoryear{Patterson, Dayton, and Graubard}{Patterson
  et~al.}{2002}]{patterson2002latent}
Patterson, B.~H., Dayton, C.~M., and Graubard, B.~I. (2002).
\newblock Latent class analysis of complex sample survey data: application to
  dietary data.
\newblock {\em Journal of the American Statistical Association} {\bf 97,}
  721--741.

\bibitem[\protect\citeauthoryear{Pfeffermann}{Pfeffermann}{1996}]{pfeffermann1996use}
Pfeffermann, D. (1996).
\newblock The use of sampling weights for survey data analysis.
\newblock {\em Statistical Methods in Medical Research} {\bf 5,} 239--261.

\bibitem[\protect\citeauthoryear{{R Core Team}}{{R Core
  Team}}{2023}]{RCoreTeam2023R}
{R Core Team} (2023).
\newblock {\em R: A Language and Environment for Statistical Computing}.
\newblock R Foundation for Statistical Computing, Vienna, Austria.

\bibitem[\protect\citeauthoryear{Sacks, Svetkey, Vollmer, Appel, Bray, Harsha,
  et~al\mbox{.}}{Sacks et~al.}{2001}]{sacks2001effects}
Sacks, F.~M., Svetkey, L.~P., Vollmer, W.~M., Appel, L.~J., Bray, G.~A.,
  Harsha, D., et~al. (2001).
\newblock Effects on blood pressure of reduced dietary sodium and the dietary
  approaches to stop hypertension (dash) diet.
\newblock {\em New England journal of medicine} {\bf 344,} 3--10.

\bibitem[\protect\citeauthoryear{Savitsky and Toth}{Savitsky and
  Toth}{2016}]{savitsky2016bayesian}
Savitsky, T.~D. and Toth, D. (2016).
\newblock Bayesian estimation under informative sampling.
\newblock {\em Electronic Journal of Statistics} {\bf 10,} 1677--1708.

\bibitem[\protect\citeauthoryear{Sotres-Alvarez, Herring, and
  Siega-Riz}{Sotres-Alvarez et~al.}{2010}]{sotres2010latent}
Sotres-Alvarez, D., Herring, A.~H., and Siega-Riz, A.~M. (2010).
\newblock Latent class analysis is useful to classify pregnant women into
  dietary patterns.
\newblock {\em The Journal of Nutrition} {\bf 140,} 2253--2259.

\bibitem[\protect\citeauthoryear{Sotres-Alvarez, Siega-Riz, Herring,
  Carmichael, Feldkamp, Hobbs, et~al\mbox{.}}{Sotres-Alvarez
  et~al.}{2013}]{sotres2013maternal}
Sotres-Alvarez, D., Siega-Riz, A.~M., Herring, A.~H., Carmichael, S.~L.,
  Feldkamp, M.~L., Hobbs, C.~A., et~al. (2013).
\newblock Maternal dietary patterns are associated with risk of neural tube and
  congenital heart defects.
\newblock {\em American Journal of Epidemiology} page kws349.

\bibitem[\protect\citeauthoryear{Spencer}{Spencer}{2000}]{spencer2000approximate}
Spencer, B.~D. (2000).
\newblock An approximate design effect for unequal weighting when measurements
  may correlate with selection probabilities.
\newblock {\em Survey Methodology} {\bf 26,} 137--138.

\bibitem[\protect\citeauthoryear{Stephens}{Stephens}{2000}]{stephens2000dealing}
Stephens, M. (2000).
\newblock Dealing with label switching in mixture models.
\newblock {\em Journal of the Royal Statistical Society Series B: Statistical
  Methodology} {\bf 62,} 795--809.

\bibitem[\protect\citeauthoryear{Stephenson, Herring, and Olshan}{Stephenson
  et~al.}{2022}]{stephenson2022derivation}
Stephenson, B.~J., Herring, A.~H., and Olshan, A.~F. (2022).
\newblock Derivation of maternal dietary patterns accounting for regional
  heterogeneity.
\newblock {\em Journal of the Royal Statistical Society Series C: Applied
  Statistics} {\bf 71,} 1957--1977.

\bibitem[\protect\citeauthoryear{Stephenson, Wu, and Dominici}{Stephenson
  et~al.}{2024}]{stephenson2024identifying}
Stephenson, B.~J., Wu, S.~M., and Dominici, F. (2024).
\newblock Identifying dietary consumption patterns from survey data: a bayesian
  nonparametric latent class model.
\newblock {\em Journal of the Royal Statistical Society Series A: Statistics in
  Society} {\bf 187,} 496--512.

\bibitem[\protect\citeauthoryear{Stephenson and Willett}{Stephenson and
  Willett}{2023}]{stephenson2023racial}
Stephenson, B. J.~K. and Willett, W.~C. (2023).
\newblock Racial and ethnic heterogeneity in diets of low-income adult females
  in the united states: results from national health and nutrition examination
  surveys from 2011 to 2018.
\newblock {\em The American Journal of Clinical Nutrition} {\bf 117,} 625--634.

\bibitem[\protect\citeauthoryear{Van~Havre, White, Rousseau, and
  Mengersen}{Van~Havre et~al.}{2015}]{van2015overfitting}
Van~Havre, Z., White, N., Rousseau, J., and Mengersen, K. (2015).
\newblock Overfitting bayesian mixture models with an unknown number of
  components.
\newblock {\em PloS One} {\bf 10,}.

\bibitem[\protect\citeauthoryear{Whelton, Carey, Aronow, Casey, Collins,
  Dennison~Himmelfarb, et~al\mbox{.}}{Whelton et~al.}{2018}]{whelton20182017}
Whelton, P.~K., Carey, R.~M., Aronow, W.~S., Casey, D.~E., Collins, K.~J.,
  Dennison~Himmelfarb, C., et~al. (2018).
\newblock 2017 acc/aha/aapa/abc/acpm/ags/apha/ash/aspc/nma/pcna guideline for
  the prevention, detection, evaluation, and management of high blood pressure
  in adults: a report of the american college of cardiology/american heart
  association task force on clinical practice guidelines.
\newblock {\em Journal of the American College of Cardiology} {\bf 71,}
  e127--e248.

\bibitem[\protect\citeauthoryear{Williams and Savitsky}{Williams and
  Savitsky}{2020}]{williams2020bayesian}
Williams, M.~R. and Savitsky, T.~D. (2020).
\newblock Bayesian estimation under informative sampling with unattenuated
  dependence.
\newblock {\em Bayesian Analysis} {\bf 15,} 57--77.

\bibitem[\protect\citeauthoryear{Williams and Savitsky}{Williams and
  Savitsky}{2021}]{williams2021uncertainty}
Williams, M.~R. and Savitsky, T.~D. (2021).
\newblock Uncertainty estimation for pseudo-bayesian inference under complex
  sampling.
\newblock {\em International Statistical Review} {\bf 89,} 72--107.

\bibitem[\protect\citeauthoryear{Zhang, Liu, Rehm, Wilde, Mande, and
  Mozaffarian}{Zhang et~al.}{2018}]{zhang2018trends}
Zhang, F.~F., Liu, J., Rehm, C.~D., Wilde, P., Mande, J.~R., and Mozaffarian,
  D. (2018).
\newblock Trends and disparities in diet quality among us adults by
  supplemental nutrition assistance program participation status.
\newblock {\em JAMA Network Open} {\bf 1,} e180237--e180237.

\end{thebibliography}


\begin{thebibliography}{}

\bibitem[\protect\citeauthoryear{Bowman, Clemens, Friday, and Moshfegh}{Bowman
  et~al.}{2020}]{bowman2020food}
Bowman, S., Clemens, J., Friday, J., and Moshfegh, A. (2020).
\newblock Food patterns equivalents database 2017--2018: methodology and user
  guide.
\newblock {\em Food Surveys Research Group: Beltsville, MD} .

\bibitem[\protect\citeauthoryear{Bowman, Clemens, Shimizu, Friday, and
  Moshfegh}{Bowman et~al.}{2018}]{bowman2018food}
Bowman, S., Clemens, J., Shimizu, M., Friday, J., and Moshfegh, A. (2018).
\newblock Food patterns equivalents database 2015--2016: methodology and user
  guide.
\newblock {\em US Department of Agriculture} .

\bibitem[\protect\citeauthoryear{Cario and Nelson}{Cario and
  Nelson}{1997}]{cario1997modeling}
Cario, M.~C. and Nelson, B.~L. (1997).
\newblock Modeling and generating random vectors with arbitrary marginal
  distributions and correlation matrix.
\newblock Technical report, Northwestern University, Department of Industrial
  Engineering and Management.

\bibitem[\protect\citeauthoryear{{Centers for Disease Control and
  Prevention}}{{Centers for Disease Control and
  Prevention}}{2022}]{centers2022how}
{Centers for Disease Control and Prevention} (2022).
\newblock How much physical activity do adults need?

\bibitem[\protect\citeauthoryear{{Dietary Guidelines Advisory
  Committee}}{{Dietary Guidelines Advisory
  Committee}}{2015}]{dietary2015dietary}
{Dietary Guidelines Advisory Committee} (2015).
\newblock {\em Dietary Guidelines for Americans 2015-2020}.
\newblock Government Printing Office.

\bibitem[\protect\citeauthoryear{Liu, Shih, Strawderman, Zhang, Johnson, and
  Chai}{Liu et~al.}{2019}]{liu2019statistical}
Liu, L., Shih, Y.-C.~T., Strawderman, R.~L., Zhang, D., Johnson, B.~A., and
  Chai, H. (2019).
\newblock Statistical analysis of zero-inflated nonnegative continuous data.
\newblock {\em Statistical Science} {\bf 34,} 253--279.

\bibitem[\protect\citeauthoryear{{National Center for Health
  Statistics}}{{National Center for Health Statistics}}{2017}]{nchs2017adult}
{National Center for Health Statistics} (2017).
\newblock Nhis - adult tobacco use - glossary.

\bibitem[\protect\citeauthoryear{Sotres-Alvarez, Siega-Riz, Herring,
  Carmichael, Feldkamp, Hobbs, et~al\mbox{.}}{Sotres-Alvarez
  et~al.}{2013}]{sotres2013maternal}
Sotres-Alvarez, D., Siega-Riz, A.~M., Herring, A.~H., Carmichael, S.~L.,
  Feldkamp, M.~L., Hobbs, C.~A., et~al. (2013).
\newblock Maternal dietary patterns are associated with risk of neural tube and
  congenital heart defects.
\newblock {\em American Journal of Epidemiology} page kws349.

\bibitem[\protect\citeauthoryear{Touloumis}{Touloumis}{2016}]{touloumis2016simulating}
Touloumis, A. (2016).
\newblock Simulating correlated binary and multinomial responses under marginal
  model specification: The simcormultres package.
\newblock {\em The R Journal} {\bf 8,} 79--91.
\newblock R package version 1.8.0.

\bibitem[\protect\citeauthoryear{Whelton, Carey, Aronow, Casey, Collins,
  Dennison~Himmelfarb, et~al\mbox{.}}{Whelton et~al.}{2018}]{whelton20182017}
Whelton, P.~K., Carey, R.~M., Aronow, W.~S., Casey, D.~E., Collins, K.~J.,
  Dennison~Himmelfarb, C., et~al. (2018).
\newblock 2017 acc/aha/aapa/abc/acpm/ags/apha/ash/aspc/nma/pcna guideline for
  the prevention, detection, evaluation, and management of high blood pressure
  in adults: a report of the american college of cardiology/american heart
  association task force on clinical practice guidelines.
\newblock {\em Journal of the American College of Cardiology} {\bf 71,}
  e127--e248.

\end{thebibliography}


\end{document}


\emergencystretch 3em
\renewcommand{\tablename}{Web Table} 
\renewcommand{\figurename}{Web Figure} 

\title{Supplementary Materials for ``Derivation of outcome-dependent dietary patterns for low-income women obtained from survey data using a Supervised Weighted Overfitted Latent Class Analysis'' by S.M. Wu, M.R. Williams, T.D. Savitsky, and B.J.K. Stephenson}

\author{Stephanie M. Wu, Matthew R. Williams, Terrance D. Savitsky, and
Briana J.K. Stephenson}

\date{}

\maketitle

\tableofcontents

\maketitle

\clearpage
\section{Web Appendix A: Gibbs Sampling Full Conditional Distributions}\label{app:gibbs}
The complete-data weighted pseudo-posterior for SWOLCA is given by
\begin{align*}\label{eq:solca}
    \tilde{p}(\bmath{\Theta}|\bmath{D}) &\propto p(\bmath{\Theta})\prod_{i=1}^n  \bigg[\pi_{c_i} \prod_{j=1}^J \prod_{r=1}^{R_j} \theta_{jc_ir}^{I(x_{ij}=r)}\frac{1}{\sqrt{2\pi}}e^{-\frac{1}{2}(z_i - \bfv_i^\intercal \bfxi_{c_i\cdot})^2}\Big\{y_iI(z_i > 0) + (1 - y_i)I(z_i \leq 0) \Big\}\bigg]^{\frac{w_i}{\kappa}}\\
    &= p(\bmath{\Theta})\prod_{i=1}^n\prod_{k=1}^K \bigg[\pi_k \prod_{j=1}^J \prod_{r=1}^{R_j} \theta_{jkr}^{I(x_{ij}=r)}\frac{1}{\sqrt{2\pi}}e^{-\frac{1}{2}(z_i - \bfv_i^\intercal \bfxi_{k\cdot})^2}\\
    &\quad \times \Big\{y_iI(z_i > 0) + (1 - y_i)I(z_i \leq 0) \Big\}\bigg]^{I(c_i=k)\frac{w_i}{\kappa}}.
\end{align*}
Priors:
\begin{align*}
    \bfpi &\sim \text{Dir}(\alpha_1, \ldots, \alpha_K)\\
    \bftheta_{jk\cdot} &\sim Dir(\eta_1, \ldots, \eta_{R_j}) \\
    \bfxi_{k\cdot} &\sim MVN_{q}(\mu_0, \Sigma_0)\\
    \mu_0&\sim N(0,1), \\
    \Sigma_0 &= \text{diag}(\sigma^2_{1},\ldots, \sigma^2_q), \ \sigma_{p}^2 \overset{iid}{\sim} \text{InvGamma}(5/2,2/5), \ p = 1,\ldots, q.
\end{align*}
We update the Gibbs sampler by iteratively sampling from the full conditional posteriors:
\begin{enumerate}
    \item \underline{Class membership probabilities, $\bfpi$}\\
    Assume a conjugate Dir($\alpha_1, \ldots, \alpha_K$) prior.
    \begin{align*}
        \tilde{p}(\bfpi|\cdot) 
        &\propto 
        \Bigg(\prod_{k=1}^K \pi_k^{\alpha_k-1}\Bigg)\Bigg\{\prod_{i=1}^n\prod_{k=1}^K\pi_{k}^{I(c_i=k)\frac{w_i}{\kappa}}\Bigg\}\\
        &= \prod_{k=1}^K \pi_k^{\alpha_k+\sum\limits_{i=1}^n \big\{I(c_i=k)\frac{w_i}{\kappa}\big\}-1}\\
        &\propto \text{Dir}\Bigg\{\alpha_1+\sum_{i=1}^nI(c_i=1)\frac{w_i}{\kappa}, \ldots, \alpha_K+\sum_{i=1}^n I(c_i=K)\frac{w_i}{\kappa}\Bigg\}.
    \end{align*}
    \item \noindent\underline{Class-specific item consumption probabilities, $\bftheta_{jk}$}\\
    Assume a conjugate Dir($\eta_1, \ldots, \eta_{R_j}$) prior.
    \begin{align*}
        \tilde{p}(\bftheta_{jk\cdot}|\cdot) 
        &\propto 
        \Bigg(\prod_{r=1}^{R_j} \theta_{jkr}^{\eta_r-1}\Bigg)\Bigg\{\prod_{i=1}^n\prod_{r=1}^{R_j} \prod_{k=1}^K\theta_{jkr}^{I(x_{ij}=r)I(c_i = k)\frac{w_i}{\kappa}}\Bigg\}, \quad j = 1,\ldots, J, \ k = 1,\ldots, K\\
        &= \prod_{r=1}^{R_j} \theta_{jkr}^{\eta_k + \sum\limits_{i=1}^n \big\{I(x_{ij}=r)I(c_i=k)\frac{w_i}{\kappa}\big\}-1}\\
        &\propto \text{Dir}\Bigg\{\eta_1+\sum_{i=1}^nI(x_{ij}=1)I(c_i=k)\frac{w_i}{\kappa}, \ldots, \eta_{R_j}+\sum_{i=1}^n I(x_{ij}={R_j})I(c_i=k)\frac{w_i}{\kappa}\Bigg\}.
    \end{align*}
    \item \noindent\underline{Class-specific probit regression parameters, $\bfxi_{k\cdot}$}\\ 
    Let $\bfz = (z_1,\ldots, z_n)^\intercal$, $V = (\bfv_1^\intercal, \ldots, \bfv_n^\intercal)^\intercal$ denote the $n\times q$ matrix of covariates, $C_k = \text{diag}\{I(c_1=k,\ldots, c_n = k)\}$ denote the $n \times n$ diagonal matrix subsetting to class $k$, and $\tilde W = \text{diag}\{w_1/\kappa, \ldots, w_n/\kappa\}$ denote the $n \times n$ diagonal matrix of normalized weights. Assume a conjugate $N_{q\times q}(\mu_0, \Sigma_0)$ prior with hyperpriors $\mu_0\sim N(0,1)$ and $\Sigma_0 = \text{diag}(\sigma^2_{1},\ldots, \sigma^2_q)$, $\sigma_{p}^2 \overset{iid}{\sim} \text{InvGamma}(5/2,2/5)$, $p=1,\ldots,q$. 
    \begin{align*}
        \tilde{p}(\bfxi_{k\cdot}|\cdot) 
        &\propto \bigg[e^{-\frac{1}{2}\left\{(\bfxi_{k\cdot}-\mu_0)^\intercal \Sigma_0^{-1}(\bfxi_{k\cdot}-\mu_0)\right\}}\bigg]\Bigg\{\prod_{i=1}^n \prod_{k=1}^K e^{-\frac{1}{2}(z_i - \bfv_i^\intercal \bfxi_{k\cdot})^2 I(c_i=k) \frac{w_i}{\kappa}}\Bigg\}, \quad k = 1,\ldots, K\\
        &= e^{-\frac{1}{2}\big\{(\bfxi_{k\cdot}-\mu_0)^\intercal \Sigma_0^{-1}(\bfxi_{k\cdot}-\mu_0)\big\}} e^{-\frac{1}{2}\big\{(\bfz-V\bfxi)^\intercal C_k \tilde{W}(\bfz-V\bfxi)\big\}}\\
        &\propto e^{-\frac{1}{2}\big(\bfxi_{k\cdot}^\intercal\Sigma_0^{-1}\bfxi_{k\cdot} -2\bfxi_{k\cdot}^\intercal\Sigma_0^{-1} \mu_0 + \bfxi_{k\cdot}^\intercal V^\intercal C_k\tilde{W}V\bfxi_{k\cdot} -2\bfxi_{k\cdot}^\intercal V^\intercal C_k \tilde{W}\bfz\big)}\\
        &= e^{-\frac{1}{2}\Big\{\bfxi_{k\cdot}^\intercal\big(\Sigma_0^{-1} + V^\intercal C_k \tilde{W}V\big)\bfxi_{k\cdot} -2\bfxi_{k\cdot}^\intercal\big(\Sigma_0^{-1} \mu_0 + V^\intercal C_k \tilde{W}\bfz\big)\Big\}}\\ 
        &\propto N\Bigg\{(\Sigma_0^{-1}+V^\intercal C_k \tilde{W} V)^{-1}(\Sigma_0^{-1}\mu_0+V^\intercal C_k \tilde{W} \bfz), (\Sigma_0^{-1}+V^\intercal\tilde{W} V)^{-1}\Bigg\}.
    \end{align*}
    \item \noindent\underline{Latent class membership assignment, $c_i$, and latent probit variable, $z_i$}\\
    \begin{align*}
        c_i |\cdot &\sim \text{Mult}\Bigg\{1, p(c_i=1|\cdot),\ldots, p(c_i=K|\cdot)\Bigg\},\\
        \text{where }&\\
        p(c_i=k|\cdot) &= \frac{p(\bfx_i, y_i, z_i, c_i=k| \bfv_i, \bmath{\Theta})}{p(\bfx_i, y_i, z_i|\bfv_i, \bmath{\Theta})}\\
        &=\frac{\pi_k \prod\limits_{j=1}^J\prod\limits_{r=1}^{R_j}\theta_{jkr}^{I(x_{ij}=r)}\frac{1}{\sqrt{2\pi}}e^{-\frac{1}{2}(z_i-\bfv_i^\intercal \bfxi_{k\cdot})^2}\Big\{y_iI(z_i>0)+(1-y_i)I(z_i\leq 0)\Big\}}{\sum\limits_{h=1}^H \Bigg[\pi_h \prod\limits_{j=1}^J\prod\limits_{r=1}^{R_j}\theta_{jhr}^{I(x_{ij}=r)}\frac{1}{\sqrt{2\pi}}e^{-\frac{1}{2}(z_i-\bfv_i^\intercal \bfxi_{h\cdot})^2}\Big\{y_iI(z_i>0)+(1-y_i)I(z_i\leq 0)\Big\}\Bigg]}.\\
        p(z_i|\cdot) 
        &\propto \frac{1}{\sqrt{2\pi}}e^{-\frac{1}{2}(z_i - \bfv_{i}^\intercal \bfxi_{c_i\cdot})^2}\Big\{y_iI(z_i > 0) + (1 - y_i)I(z_i \leq 0)\Big\}\\
        \implies z_i|\cdot &\sim \begin{cases}
        \text{TruncNormal}(\bfv_i^\intercal \bfxi_{c_i\cdot}, 1, 0, \infty)& \text{if }y_i=1\\
            \text{TruncNormal}(\bfv_i^\intercal \bfxi_{c_i\cdot}, 1, -\infty, 0)& \text{if }y_i=0.
        \end{cases} 
    \end{align*}
\end{enumerate}

\clearpage
\section{Web Appendix B: Detailed Simulation Methods}
\subsection{Simulation Design}
Data are generated for a finite population of size $N = 80,000$. A total of $K_{true}=3$ latent classes (i.e., dietary patterns) exist in the population, with probability of membership distributed as $\bmath{\pi}=(0.575, 0.250, 0.175)^T$. The population is comprised of two strata of unequal sizes, with stratum 1 having 20,000 individuals and stratum 2 having 60,000 individuals. Each individual is randomly assigned to a stratum, then to one of the three latent classes according to stratum-specific probabilities, equal to $(0.2, 0.4, 0.4)^T$ for stratum 1 and $(0.7, 0.2, 0.1)^T$ for stratum 2. Since the class membership probabilities are stratum-dependent, there is correlation between latent class and selection into the sample. Although membership probabilities are indexed by stratum in the population, our inferential interest is in the population average latent class membership probabilities, as data analysts may only have access to published weights and not strata memberships because the weights include unit-level non-response adjustments.

The latent classes correspond to three dietary patterns. Each pattern consists of $J=30$ categorical food items, each consumed at one of $R_j=R=4$ levels: none, low, medium, and high. Pattern 1 is defined as no consumption for the first 15 items and medium consumption for the last 15; Pattern 2 as high for the first 6 items and low for the remaining 24; and Pattern 3 as medium for the first 9 items, high for the next 12, and level 1 for the remaining 9 (left-hand-side of Web Figure~\ref{f:sim_theta}). For each individual $i \in \{1,\ldots,n\}$ and food item $j \in \{1,\ldots J\}$, the categorical exposure variable $x_{ij}$ is drawn from a Categorical($\theta_{jc_i1}, \ldots, \theta_{jc_iR}$) distribution, where $c_i \in \{1,\ldots, K_{true}\}$ is the individual's pattern and $\theta_{jc_ir}$, $r \in\{1, \ldots, R\}$, is 0.85 for the modal consumption level and 0.05 for the remaining three levels. 

The binary outcome is drawn from a probit model with coefficient values $\bmath{\xi} = (1, 0.3, -0.5, 0.5,$ $-0.7, -1.3)^T$, corresponding to approximate hypertension risk of 0.84, 0.62, and 0.31 for the three respective latent classes in stratum 1, and 0.69, 0.24, and 0.1 for the latent classes in stratum 2. Notice that hypertension risk decreases from latent class 1 to 2 to 3 and is also lower for those in stratum 2 than in stratum 1. Informative sampling is present due to the correlation between stratum and outcome, and the effect of latent class on the outcome is modified by stratum. We also introduce clustering in the outcome to mimic correlated outcomes among areas that may be sampled together. Each cluster is composed of 50 individuals, with 400 clusters in stratum 1 and 1200 clusters in stratum 2. R package \texttt{SimCorMultRes} version 1.8.0 \citep{touloumis2016simulating} is used to create correlated binary outcomes within clusters using the modified NORmal To Anything (NORTA) method \citep{cario1997modeling}, assuming an exchangeable latent correlation matrix with correlation 0.5 on the off-diagonals. 

Models are initiated with $K=30$ and Dirichlet hyperparameter  $\alpha = 1/K$ for all $k \in \{1\ldots, K\}$ to encourage sparsity and moderate growth of new pattern formation. A noninformative flat Dir(1) prior is used for $\bmath{\theta}_{jc_i\cdot}$, and weakly informative priors are used for the regression parameters $\bmath{\xi}$. The default sample setting involves sampling 5\% of the population ($n=4000$) using a stratified sampling scheme with unequal sampling probabilities where 2000 individuals are sampled from each stratum. Individuals sampled from stratum 1 have a sampling weight of 10, and individuals sampled from stratum 2 have a sampling weight of 30. 

We examine model performance and sensitivity under nine sampling and data-generating scenarios, enumerated below. We examine three \textit{survey sampling designs}: simple random sampling (SRS) with no systematic bias or sample-population differences; stratified sampling with unequal sampling probabilities; or stratified cluster sampling with unequal sampling probabilities and correlated outcomes. We assess three \textit{associations of interest}: a conditional outcome model with stratum included as a covariate; a marginal outcome model that does not condition on selection or adjust for selection bias; or a model with two additional covariates, one binary and one continuous. We compare the models under three different \textit{sample sizes}: 1\% of the population (n = 800); 5\% of the population (n = 4000); or 10\% of the population (n = 8000). We inspect model robustness to three \textit{pattern settings}: strongly identified patterns with mode 85\% for the true consumption level and 5\% for the remaining non-modal levels; weakly identified patterns with mode 55\% for the true consumption level and 15\% otherwise; and overlap patterns where two of the patterns are identical except for a few items that drive differences in the outcome (see Web Figure~\ref{f:sim_theta}). Bold text indicates deviation from the default setting of stratified sampling, conditional model, sample size 4000, and mode 85\%. 
\begin{enumerate}
    \item \textbf{SRS}, conditional, n=4000, mode 85\%
    \item \textbf{Stratified sampling}, conditional, n=4000, mode 85\%
    \item \textbf{Stratified cluster sampling}, conditional, n=4000, mode 85\%
    \item Stratified sampling, \textbf{marginal}, n=4000, mode 85\%
    \item Stratified sampling, \textbf{additional}, n=4000, mode 85\%
    \item Stratified sampling, conditional, \textbf{n=8000}, mode 85\%
    \item Stratified sampling, conditional, \textbf{n=800}, mode 85\%
    \item Stratified sampling, conditional, n=4000, \textbf{mode 55\%}
    \item Stratified sampling, conditional, n=4000, \textbf{overlap}
\end{enumerate}

We obtain results for each simulation scenario over 100 independent iterations characterized by different observed sample data. To compare model performance for parameter estimation and inference, we use the measures described below, with calculations for the parameter $\pi$ provided as an example.
\begin{enumerate}
    \item \textbf{Mean absolute bias:} Mean absolute bias was measured as the mean absolute distance between the estimated and true parameter values, averaged across $L=100$ iterations: 
    $\frac{1}{LK}\sum_{l,k} |\pi^{(l)}_k - \hat{\pi}^{(l)}_k|$, where $l=1,\ldots,100$ denotes the simulation iteration and $k=1,\ldots, K$ denotes the latent class.
    \item \textbf{Variability:} Variability was measured as the full width of the 95\% quantile credible interval from posterior samples, averaged across latent classes and simulation iterations: $\frac{1}{LK}\sum_{l,k}(\hat{\pi}^{(l)}_{k, 0.975} - \hat{\pi}^{(l)}_{k, 0.025})$, where $\hat{\pi}^{(l)}_{k, q}$ denotes the $q$-th quantile of the estimated $\pi$ value for simulation iteration $l$ and latent class $k$.
    \item \textbf{Coverage:} Coverage was measured as the proportion of the 95\% quantile credible intervals over the 100 iterations that covered the true parameter values in the population, averaged across latent classes: $\frac{1}{LK}\sum_{lk} I[\hat{\pi}^{(l)}_{k, 0.025} \leq \pi \leq \hat{\pi}^{(l)}_{k, 0.975}]$, where $I[\cdot]$ denotes the indicator function. 
\end{enumerate}
Results from all simulation settings are displayed in Web Table~\ref{t:full_sims}.

\begin{figure}[H]
    \centering
    \caption{Simulated modal item consumption levels for the default setting with disjoint patterns (a) and the overlap setting where consumption of many foods is the same for two patterns (b). \textmd{Modal consumption level is defined as $\text{argmax}_r\theta_{jkr}$ for $r=1,\ldots,4,\ j = 1,\ldots, 30,\ k = 1,\ldots, 3$.}}
    \begin{subfigure}[t]{0.41\textwidth}
        \includegraphics[width = \textwidth]{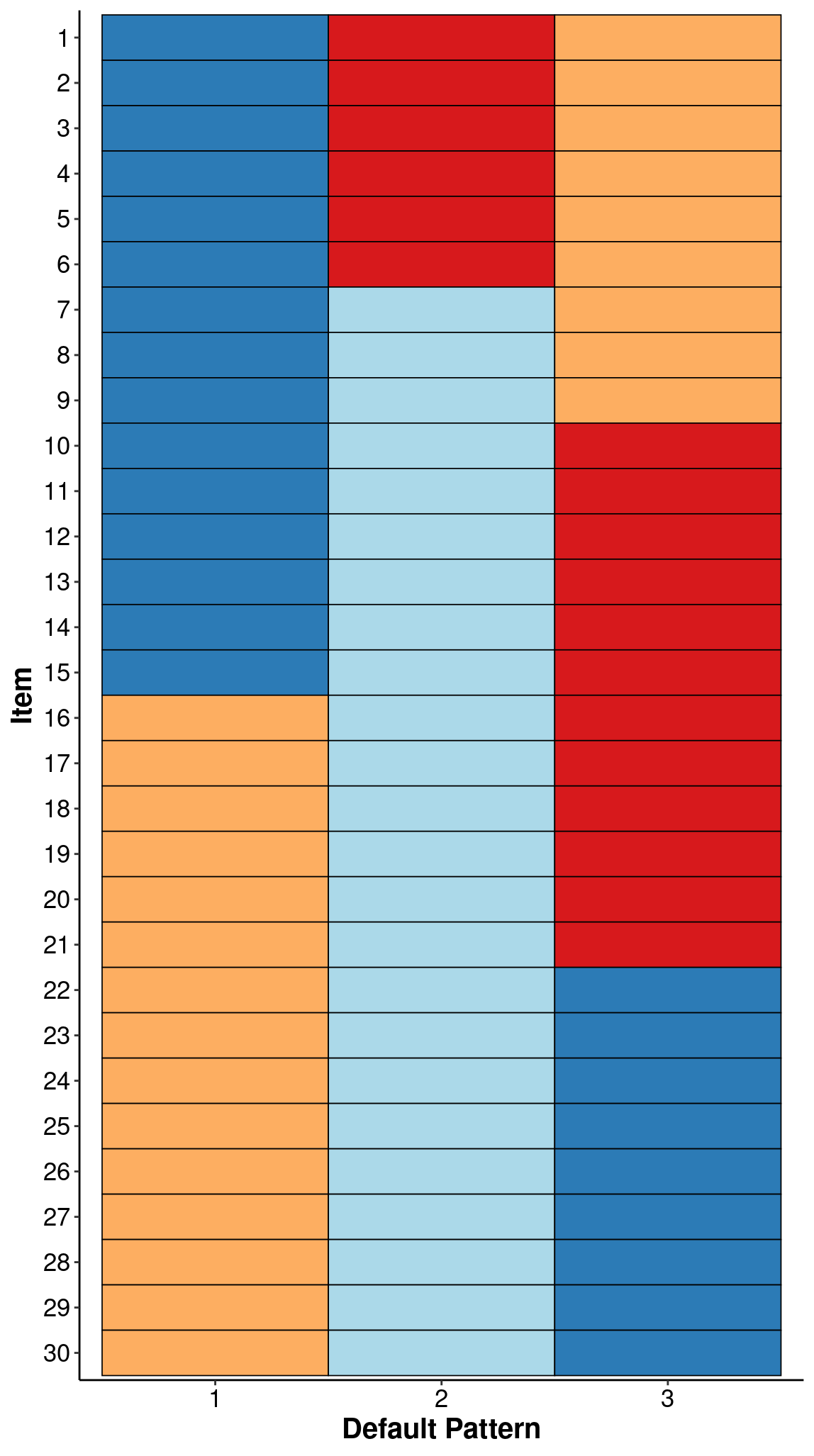}
        \caption{\textmd{Default patterns for scenarios 1 -- 8}}
    \end{subfigure}
    \begin{subfigure}[t]{0.58\textwidth}
        \includegraphics[width = 0.95\textwidth, right]{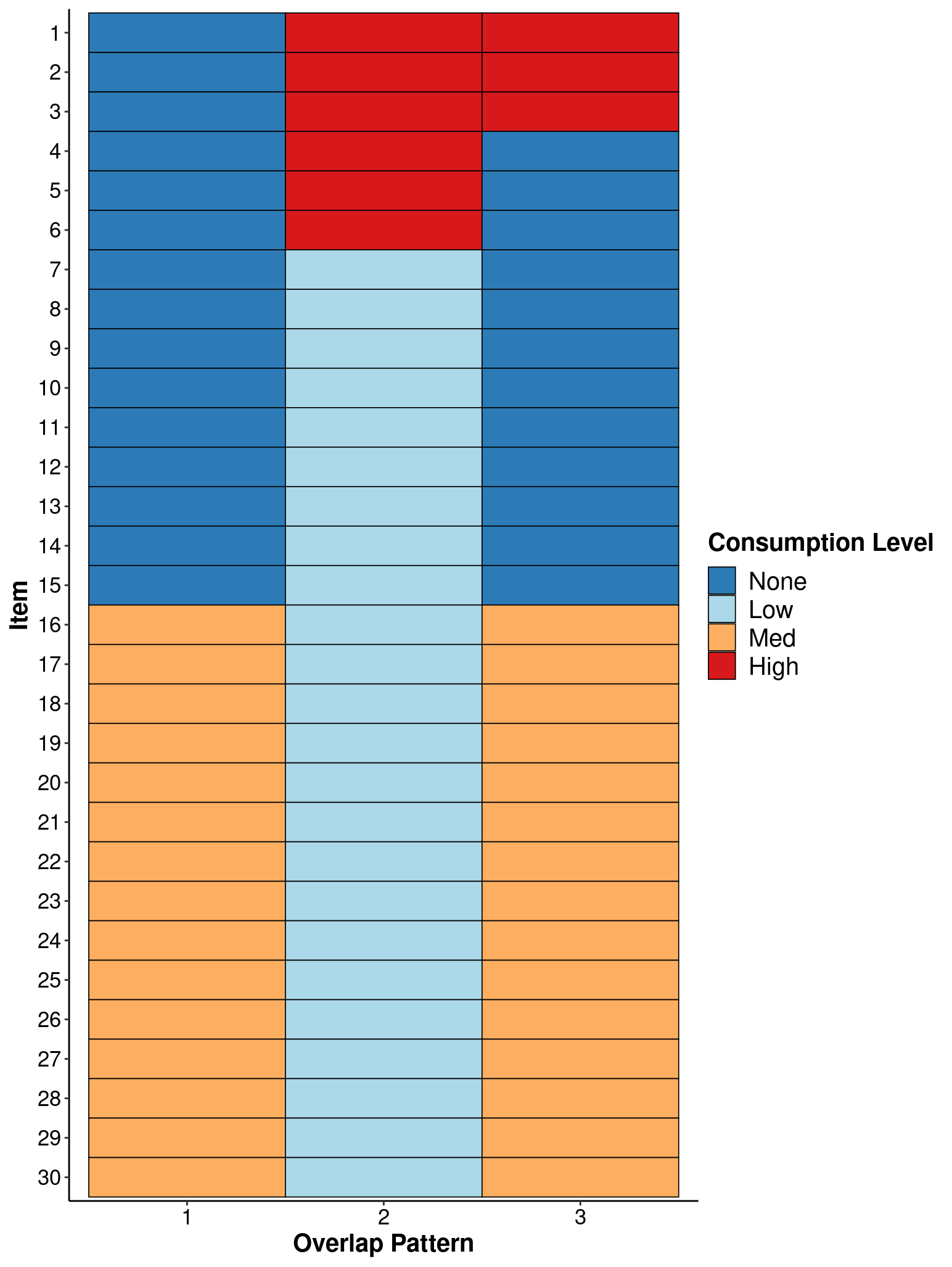}
        \caption{\textmd{Overlap patterns for scenario 9}}
    \end{subfigure}
    \label{f:sim_theta}
\end{figure}

\subsubsection{Web Figure~\ref{f:sim_theta}: Simulated Item Consumption Probabilities}

\subsection{Simulation Results}
\subsubsection{Web Table~\ref{t:full_sims}: Full Simulation Results}
\begin{table}[H]
\footnotesize
\caption{Summary of mean absolute bias, 95\% credible interval width, and coverage based on posterior MCMC samples from 100 independent draws from the population. \textmd{Models: unweighted SOLCA, two-step WOLCA, and proposed SWOLCA. Parameters: number of classes $K$, class membership probabilities $\bmath{\pi}$, item consumption level probabilities $\bftheta$, regression coefficients $\bfxi$. Strat = stratified sampling, Strat Cl = stratified cluster sampling, Cond = conditional model, Marg = marginal model, Add'l = additional covariates}}
\centering
\setlength{\tabcolsep}{0.33em}
\setlength\cellspacetoplimit{5pt}
\setlength\cellspacebottomlimit{5pt}
\begin{tabular}[t]{llcccccccccc}
\toprule
& & \multicolumn{4}{Sc}{\textbf{Absolute Bias}} & \multicolumn{3}{c}{\textbf{CI Width}} & \multicolumn{3}{c}{\textbf{Coverage}}\\
\cmidrule(lr){3-6}
\cmidrule(lr){7-9}
\cmidrule(l){10-12}
\textbf{Scenario} & \textbf{Model} & $\bmath{K}$ & $\bmath{\pi}$ & $\bmath{\theta}$ & $\bmath{\xi}$ & $\bmath{\pi}$ & $\bmath{\theta}$ & $\bmath{\xi}$ & $\bmath{\pi}$ & $\bmath{\theta}$ & $\bmath{\xi}$ \\
\midrule
(1) \textbf{SRS}, Cond, n=4000, Mode 85\% & SOLCA & 0.00 & 0.006 & 0.006 & 0.063 & 0.027 & 0.042 & 0.367 & 0.957 & 0.958 & 0.965\\
& WOLCA & 0.00 & 0.006 & 0.006 & 0.063 & 0.036 & 0.044 & 0.762 & 0.950 & 0.958 & 0.992\\
& SWOLCA & 0.00 & 0.006 & 0.006 & 0.063 & 0.027 & 0.042 & 0.419 & 0.947 & 0.953 & 0.983\\[1ex]
(2) \textbf{Strat}, Cond, n=4000, Mode 85\% & SOLCA & 0.00 & 0.081 & 0.006 & 0.047 & 0.069 & 0.045 & 0.374 & 0.190 & 0.962 & 0.972\\
& WOLCA & 0.00 & 0.006 & 0.007 & 0.043 & 0.031 & 0.045 & 0.672 & 0.957 & 0.933 & 0.998\\
& SWOLCA & 0.00 & 0.006 & 0.006 & 0.044 & 0.036 & 0.049 & 0.414 & 0.977 & 0.952 & 0.990\\[1ex]
(3) \textbf{Strat Cl}, Cond, n=4000, Mode 85\% & SOLCA & 0.00 & 0.082 & 0.006 & 0.132 & 0.074 & 0.046 & 0.390 & 0.223 & 0.966 & 0.592\\
& WOLCA & 0.00 & 0.006 & 0.006 & 0.127 & 0.037 & 0.044 & 1.210 & 0.963 & 0.942 & 0.990\\
& SWOLCA & 0.00 & 0.006 & 0.006 & 0.126 & 0.031 & 0.047 & 0.816 & 0.950 & 0.942 & 0.963\\[1ex]
(4) Strat, \textbf{Marg}, n=4000, Mode 85\% & SOLCA & 0.00 & 0.008 & 0.006 & 0.203 & 0.062 & 0.043 & 0.162 & 0.963 & 0.958 & 0.063\\
& WOLCA & 0.00 & 0.016 & 0.007 & 0.031 & 0.107 & 0.049 & 0.348 & 0.947 & 0.939 & 0.993\\
& SWOLCA & 0.00 & 0.011 & 0.007 & 0.033 & 0.097 & 0.063 & 0.278 & 0.967 & 0.965 & 0.987\\[1ex]
(5) Strat, \textbf{Add'l}, n=4000, Mode 85\% & SOLCA & 0.00 & 0.083 & 0.006 & 0.099 & 0.043 & 0.041 & 0.408 & 0.057 & 0.958 & 0.932\\
& WOLCA & 0.00 & 0.005 & 0.006 & 0.098 & 0.037 & 0.045 & 0.861 & 0.967 & 0.941 & 0.973\\
& SWOLCA & 0.04 & 0.009 & 0.010 & 0.110 & 0.037 & 0.051 & 0.456 & 0.950 & 0.924 & 0.928\\[1ex]
(6) Strat, Cond, \textbf{n=8000}, Mode 85\% & SOLCA & 0.00 & 0.080 & 0.005 & 0.049 & 0.076 & 0.042 & 0.367 & 0.227 & 0.972 & 0.980\\
& WOLCA & 0.06 & 0.010 & 0.011 & 0.038 & 0.044 & 0.044 & 0.519 & 0.920 & 0.908 & 0.960\\
& SWOLCA & 0.00 & 0.004 & 0.005 & 0.030 & 0.029 & 0.038 & 0.373 & 0.967 & 0.953 & 0.997\\[1ex]
(7) Strat, Cond, \textbf{n=800}, Mode 85\% & SOLCA & 0.00 & 0.084 & 0.013 & 0.098 & 0.064 & 0.088 & 0.701 & 0.027 & 0.938 & 0.945\\
& WOLCA & 0.00 & 0.013 & 0.014 & 0.099 & 0.060 & 0.095 & 1.371 & 0.933 & 0.919 & 0.983\\
& SWOLCA & 0.00 & 0.013 & 0.014 & 0.097 & 0.062 & 0.099 & 0.687 & 0.947 & 0.922 & 0.947\\[1ex]
(8) Strat, Cond, n=4000, \textbf{Mode 55\%} & SOLCA & 0.03 & 0.123 & 0.009 & 0.049 & 0.137 & 0.077 & 0.455 & 0.500 & 0.977 & 0.960\\
& WOLCA & 0.01 & 0.027 & 0.012 & 0.043 & 0.127 & 0.086 & 0.668 & 0.913 & 0.972 & 0.997\\
& SWOLCA & 0.00 & 0.018 & 0.012 & 0.055 & 0.118 & 0.116 & 0.695 & 0.927 & 0.984 & 0.995\\[1ex]
(9) Strat, Cond, n=4000, \textbf{Overlap} & SOLCA & 0.00 & 0.082 & 0.006 & 0.047 & 0.029 & 0.040 & 0.306 & 0.000 & 0.955 & 0.920\\
& WOLCA & 0.00 & 0.005 & 0.006 & 0.092 & 0.029 & 0.043 & 0.620 & 0.977 & 0.940 & 0.887\\
& SWOLCA & 0.00 & 0.005 & 0.006 & 0.056 & 0.030 & 0.045 & 0.368 & 0.973 & 0.955 & 0.942\\
\bottomrule
\end{tabular}
\label{t:full_sims}
\end{table} 

\begin{figure}[H]
    \centering
    \caption{Point estimates and 95\% CIs for the probability of outcome for a single illustrative iteration under the marginal association of interest setting (simulation scenario 4). \textmd{The solid red line indicates true population values, the dashed red line indicates sample values, and $C$ indicates latent class assignment. Error bars are used because the two-step WOLCA does not contain posterior samples for the outcome regression coefficients.}}
    \includegraphics[width = 0.7\textwidth]{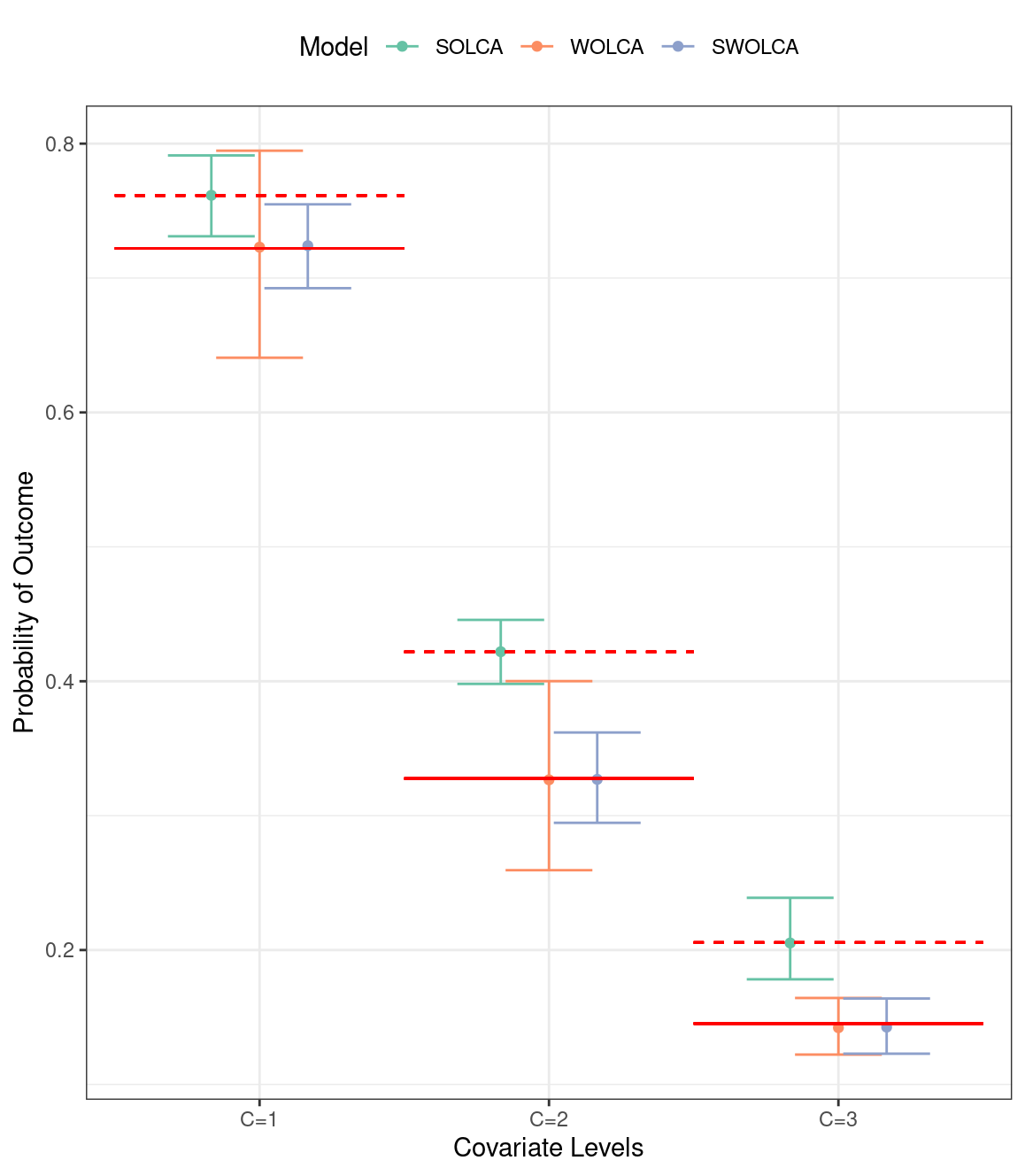}
    \label{f:sims_phi_marg}
\end{figure}

\clearpage
\subsubsection{Web Figure~\ref{f:sims_phi_marg}: Outcome Probability Estimation for Marginal Association of Interest}

\clearpage
\section{Web Appendix C: Data Application}
\subsection{Data Description}
\subsubsection{Dietary Intake Data}
Dietary intake was collected via the ``What We Eat in America'' survey component of NHANES, where two 24-hour dietary recalls recorded intake of food items and beverages. Corresponding nutrient intake was calculated using the Food and Nutrition Database for Dietary Studies (FNDDS) and converted into food item groups based on the food pattern equivalent per 100g of consumption \citep{dietary2015dietary, bowman2018food}. Dietary exposure data were summarized into 28 distinct food item groups and converted to categorical variables with four levels of consumption: none, low, medium, and high. The consumption levels were calculated using the tertiles of positive consumption for each food group \citep{liu2019statistical, sotres2013maternal, stephenson2022racial}. Web Table~\ref{t:fped} provides a list of all 28 food item groups alongside a brief description of the foods included and units used. 
\subsubsection{Hypertension Data}
Blood pressure was measured during a standard physical exam in a mobile examination center (MEC) on at least three separate occasions. Hypertension was calculated as a binary, with a positive result arising if the average systolic blood pressure was above 130 or the average diastolic blood pressure was above 80, as set by the American College of Cardiology and American Heart Association Guidelines \citep{whelton20182017}. A positive hypertension result was also given to those who reported being told they had high blood pressure at least twice, as well as those who reported currently taking prescribed medication for high blood pressure.
\subsubsection{Demographic Data}
Other major demographic risk factors for hypertension were included in the probit regression model as additional covariates. These consisted of: age (3 categories: 20-39, 40-59, $ge 60$ years old); race and ethnicity (5 categories: Non-Hispanic (NH) White, NH Black, NH Asian, Hispanic/Latino, and Mixed/Other); current smoker (binary), defined as having smoked at least 100 lifetime cigarettes and currently smoking cigarettes \citep{nchs2017adult}; and physical activity (binary), with ``active'' defined as having at least 150 minutes of moderate or vigorous exercise per week \citep{centers2022how}. Web Table~\ref{t:demog_summary} displays summaries of these demographic characteristics by hypertension outcome among sampled low-income women. Prevalence of hypertension was 53.6\% overall and was higher for those at least 65 years of age, identifying at NH Black, and physically inactive. 

\subsubsection{Web Table~\ref{t:fped}: Description of Foods Included in Food Item Groups}
\begin{table}[H]
    \centering
    \footnotesize
    \caption{Food item groups with a brief description of foods included and the units used. \textmd{Adapted from the Food Pattern Equivalents Database Methodology and User Guide \citep{bowman2020food}. Abbreviaations: eq. = equivalent, oz. = ounce, tsp. = teaspoon, num. = number of.}}
    \begin{tabular}{p{0.2\linewidth} p{0.65\linewidth} p{0.06\linewidth}}
    \toprule
    \textbf{Food Item Group} & \textbf{Food Description} & \textbf{Units}\\ \midrule
    Citrus/Melons/Berries & Intact fruits (whole or cut) of citrus, melons, and berries & cup eq.\\ 
    Other Fruits & Intact fruits (whole or cut); excluding citrus, melons, and berries & cup eq.\\
    Fruit Juice & Fruit juices, citrus and non-citrus & cup eq.\\
    Dark Green Vegs & Dark green vegetables & cup eq.\\
    Tomatoes & Tomatoes and tomato products & cup eq.\\
    Oth Red/Orange Vegs & Other red and orange vegetables, excluding tomatoes and tomato products & cup eq.\\
    Potatoes & White potatoes & cup eq.\\
    Other Starchy Vegs & Other starchy vegetables, excluding white potatoes & cup eq.\\ 
    Other Vegs & Other vegetables not in the vegetable components listed above & cup eq.\\
    Whole Grains & Grains defined as whole grains and containing the entire grain kernel: the bran, germ, and endosperm & oz. eq.\\
    Refined Grains & Refined grains that do not contain all of the components of the entire grain kernel & oz. eq.\\
    Meat & Beef, veal, pork, lamb, and game meat; excludes organ meat and cured meat & oz. eq.\\
    Cured Meats & Frankfrurters, sausages, corned beef, cured ham and luncheon meat that are made from beef, pork, or poultry & oz. eq.\\
    Organ Meat & Organ meat from beef, veal, pork, lamb, game, and poultry & oz. eq.\\
    Poultry & Chicken, turkey, Cornish hens, duck, goose, quail, and pheasant (game birds); excludes organ meat and cured meat & oz. eq.\\
    Seafood (High n-3) & Seafood (finfish, shellfish, and other seafood) high in $n$-3 fatty acids & oz. eq.\\
    Seafood (Low n-3) & Seaffod (finfish, shellfish, and other seafood) low in $n$-3 fatty acids & oz. eq.\\
    Eggs & Eggs (chicken, duck, goose, quail) and egg substitutes (oz. eq)\\ 
    Soy Products & Soy products, excluding calcium fortified soy milk (soymilk) and products made with raw (green) soybean & oz. eq.\\
    Nuts and Seeds & Peanuts, tree nuts, and seeds; excludes coconut & oz. eq.\\
    Legumes (Protein) & Beans, peas, and lentiles (legumes) computed as protein foods & oz. eq.\\
    Milk & Fluid milk, buttermilk, evaporated milk, dry milk, and calcium fortified soy milk (soymilk) & cup eq.\\
    Yogurt & Yogurt & cup eq.\\
    Cheese & Cheeses & cup eq.\\
    Oils & Fats naturally present in nuts, seeds, and seafood; all unhydrogentaed vegetable oils, except palm oil, palm kernel oil, and coconut oils; the fat present in avocado and olives above the allowable amount; 50\% of the fat present in stick and tub margarines and margarine spreads & grams\\
    Solid Fats & Fats naturally present in meat, poultry, eggs, and dairy (lard, tallow, and butter); fully or partially hydrogenated oils; shortening; palm oil; palm kernel oil; coconut oils; fats naturally present in coconut meat and cocoa butter; and 50\% of the fat present in stick and tub margarines and margarine spreads & grams\\
    Added Sugars & Caloric sweeteners such as syrups and sugars and others defined as added sugars & tsp. eq.\\
    Alcoholic Drinks & Alcoholic beverages and alcohol (ethanol) added to foods after cooking & num. drinks
    \end{tabular}
    \label{t:fped}
\end{table}

\subsubsection{Web Table~\ref{t:demog_summary}: Demographic Summary by Hypertension Outcome}
\begin{table}[H]
    \normalsize
    \centering
    \caption{Summary statistics of demographic variables selected for the SWOLCA probit model, grouped by hypertension outcome, among sampled individuals with complete data ($n = 2003$). \textmd{Column-wise percentages are shown in parentheses.}}
    \begin{tabular}{@{}lllll@{}}
    \toprule
    Variable & Level & No Hypertension & Hypertension & Total \\ \midrule
    Sample Size &  & 930 & 1073 & 2003 \\ \addlinespace
    Age Group & {[}20, 40) & 544 (58.5\%) & 120 (11.2\%) & 664 (33.2\%) \\
     & {[}40, 60) & 277 (29.8\%) & 360 (33.6\%) & 637 (31.8\%) \\
     & \textgreater{}=60 & 109 (11.7\%) & 593 (55.3\%) & 702 (35.0\%) \\ \addlinespace
    Race/Ethnicity & NH White & 270 (29.0\%) & 323 (30.1\%) & 593 (29.6\%) \\
     & NH Black & 189 (20.3\%) & 307 (28.6\%) & 496 (24.8\%) \\
     & NH Asian & 84 (9.0\%) & 65 (6.1\%) & 149 (7.4\%) \\
     & Hispanic/Latino & 352 (37.8\%) & 331 (30.8\%) & 683 (34.1\%) \\
     & Other/Mixed & 35 (3.8\%) & 47 (4.4\%) & 82 (4.1\%) \\ \addlinespace
    Current Smoker & No & 731 (78.6\%) & 846 (78.8\%) & 1577 (78.7\%) \\
     & Yes & 199 (21.4\%) & 227 (21.2\%) & 426 (21.3\%) \\ \addlinespace
    Physical Activity & Inactive & 398 (42.8\%) & 624 (58.2\%) & 1022 (51.0\%) \\
     & Active & 532 (57.2\%) & 449 (41.8\%) & 981 (49.0\%) \\ \bottomrule
    \end{tabular}
    \label{t:demog_summary}
\end{table}

\subsection{Results for the Proposed SWOLCA Model}
\subsubsection{Web Table~\ref{t:SWOLCA_coefs}: SWOLCA Full Regression Output}
\begin{table}[H]
\scriptsize
    \centering
    \caption{Full regression parameter estimates for SWOLCA, adjusting for demographic confounders. \textmd{Reference group: Multicultural diet, age [20,40), NH White, non-smoker, physically inactive. Class2 = Healthy American, Class3 = Western, Class4 = Restricted Vegetarian, Class5 = Restricted American.}}
    \begin{tabular}{l>{}c>{}c>{}c}
    \toprule
    Covariate & Median & 95\% Credible Interval & P($\xi$ > 0)\\
    \midrule
    Reference & -1.66 & (-2.92, -0.41) & $<$0.01\\
    Class2 & 0.45 & (-1.12,  1.96) & 0.70\\
    Class3 & 0.49 & (-0.86,  1.83) & 0.75\\
    Class4 & 0.67 & (-0.88,  2.17) & 0.79\\
    Class5 & 1.01 & (-0.21,  2.15) & 0.96\\
    {}[40,60) & 0.43 & (-0.24,  1.15) & 0.89\\
    $>=$60 & 2.27 & (1.35, 3.25) & 1.00\\
    NH Black & 0.89 & (-0.48,  2.20) & 0.91\\
    NH Asian & 0.62 & (-0.17,  1.39) & 0.94\\
    Hispanic/Latino & 0.38 & (-1.70,  2.43) & 0.65\\
    Other/Mixed & 1.48 & (-4.66,  7.05) & 0.69\\
    Smoker & 0.05 & (-1.63,  1.82) & 0.52\\
    Active & 0.13 & (-1.38,  1.75) & 0.56\\
    \addlinespace
    {}[40,60):Class2 & 0.74 & (-0.19,  1.70) & 0.94\\
    {}[40,60):Class3 & 0.71 & (-0.11,  1.56) & 0.95\\
    {}[40,60):Class4 & 0.91 & (-0.08,  1.92) & 0.96\\
    {}[40,60):Class5 & 0.85 & (-0.14,  1.91) & 0.95\\
    >=60:Class2 & -0.14 & (-1.99,  1.57) & 0.44\\
    >=60:Class3 & -0.13 & (-1.32,  1.02) & 0.40\\
    >=60:Class4 & -0.42 & (-1.86,  1.00) & 0.30\\
    >=60:Class5 & -0.36 & (-1.22,  0.54) & 0.22\\
    \addlinespace
    NH Black:Class2 & -0.22 & (-1.93,  1.53) & 0.40\\
    NH Black:Class3 & -0.50 & (-1.92,  1.01) & 0.24\\
    NH Black:Class4 & -0.39 & (-1.83,  1.07) & 0.30\\
    NH Black:Class5 & -0.27 & (-1.62,  1.14) & 0.35\\
    NH Asian:Class2 & -0.72 & (-5.07,  3.42) & 0.37\\
    NH Asian:Class3 & -0.47 & (-1.73,  0.71) & 0.22\\
    NH Asian:Class4 & -0.86 & (-1.98,  0.27) & 0.07\\
    NH Asian:Class5 & -0.39 & (-5.64,  5.12) & 0.45\\
    Hispanic/Latino:Class2 & -0.32 & (-2.70,  1.96) & 0.39\\
    Hispanic/Latino:Class3 & -0.47 & (-2.64,  1.64) & 0.32\\
    Hispanic/Latino:Class4 & -0.70 & (-2.89,  1.45) & 0.26\\
    Hispanic/Latino:Class5 & -0.66 & (-3.74,  2.43) & 0.35\\
    Other/Mixed:Class2 & -1.28 & (-7.31,  5.17) & 0.34\\
    Other/Mixed:Class3 & -0.47 & (-6.26,  5.88) & 0.44\\
    Other/Mixed:Class4 & -1.21 & (-6.78,  4.83) & 0.34\\
    Other/Mixed:Class5 & -1.08 & (-6.52,  4.99) & 0.35\\
    \addlinespace
    Smoker:Class2 & 0.60 & (-0.89,  2.03) & 0.78\\
    Smoker:Class3 & -0.06 & (-1.87,  1.68) & 0.48\\
    Smoker:Class4 & -0.12 & (-1.90,  1.66) & 0.45\\
    Smoker:Class5 & -0.27 & (-2.81,  2.19) & 0.41\\
    \addlinespace
    Active:Class2 & -0.43 & (-2.32,  1.41) & 0.34\\
    Active:Class3 & -0.37 & (-2.13,  1.25) & 0.33\\
    Active:Class4 & -0.14 & (-1.75,  1.45) & 0.44\\
    Active:Class5 & -0.68 & (-2.28,  0.93) & 0.18\\
    \bottomrule
    \end{tabular}
    \label{t:SWOLCA_coefs}
\end{table}

\subsection{Results for the Unweighted SOLCA Model}
\begin{figure}[H]
    \centering
    \begin{subfigure}[t]{0.34\textwidth}
        \centering
        \includegraphics[width=\textwidth]{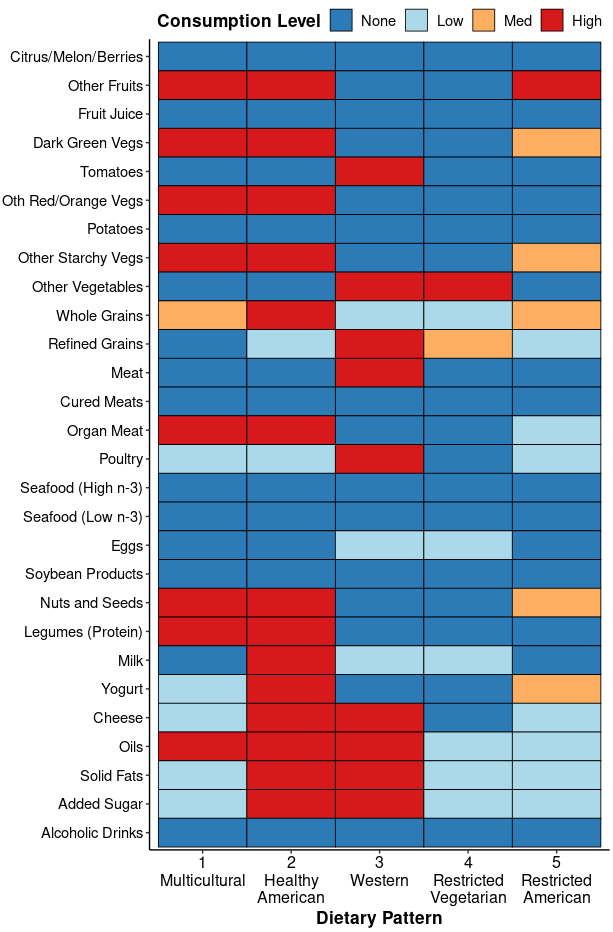}
        \caption{\textmd{Modal consumption}}
    \end{subfigure}
    \begin{subfigure}[t]{0.65\textwidth}
        \centering
        \includegraphics[width=\textwidth]{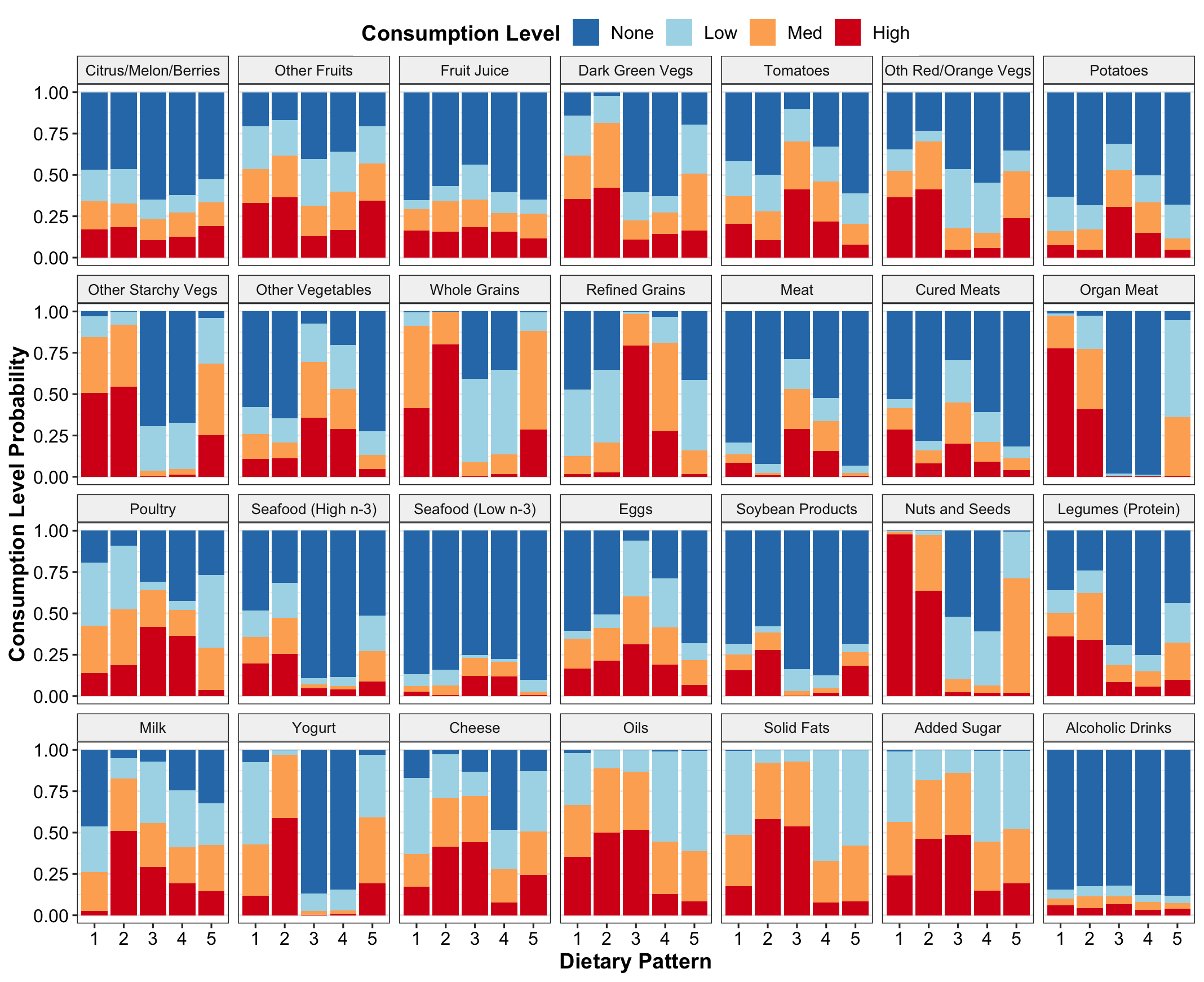}
        \caption{\textmd{Consumption level probabilities}}
    \end{subfigure}
    \caption{Five diet-hypertension patterns identified by the unweighted SOLCA model among low-income women in the US. \textmd{Consumption levels are categorized as none, low, medium, and high. (a) For each pattern, consumption of each food component is colored according to the modal consumption level (i.e., $\text{argmax}_r \theta_{jkr}$ for $r=1,\ldots, 4$, $j = 1,\ldots, 28$, $k = 1,\ldots, 5$). (b) Detailed breakdown of consumption level probabilities by diet-hypertension pattern for each food component.}}
    \label{f:SOLCA_theta}
\end{figure}

\subsubsection{Web Figure~\ref{f:SOLCA_theta}: SOLCA Diet-Hypertension Patterns}

\begin{figure}[H]
    \centering
    \caption{Estimated probability of hypertension outcome by diet-hypertension pattern for all covariates, including interactions with pattern, for the unweighted SOLCA model. \textmd{For each covariate plot, all other covariates are set to the following baseline values: Multicultural diet, age [20,40), NH White, non-smoker, and physically inactive.}}
    \includegraphics[width = \linewidth]{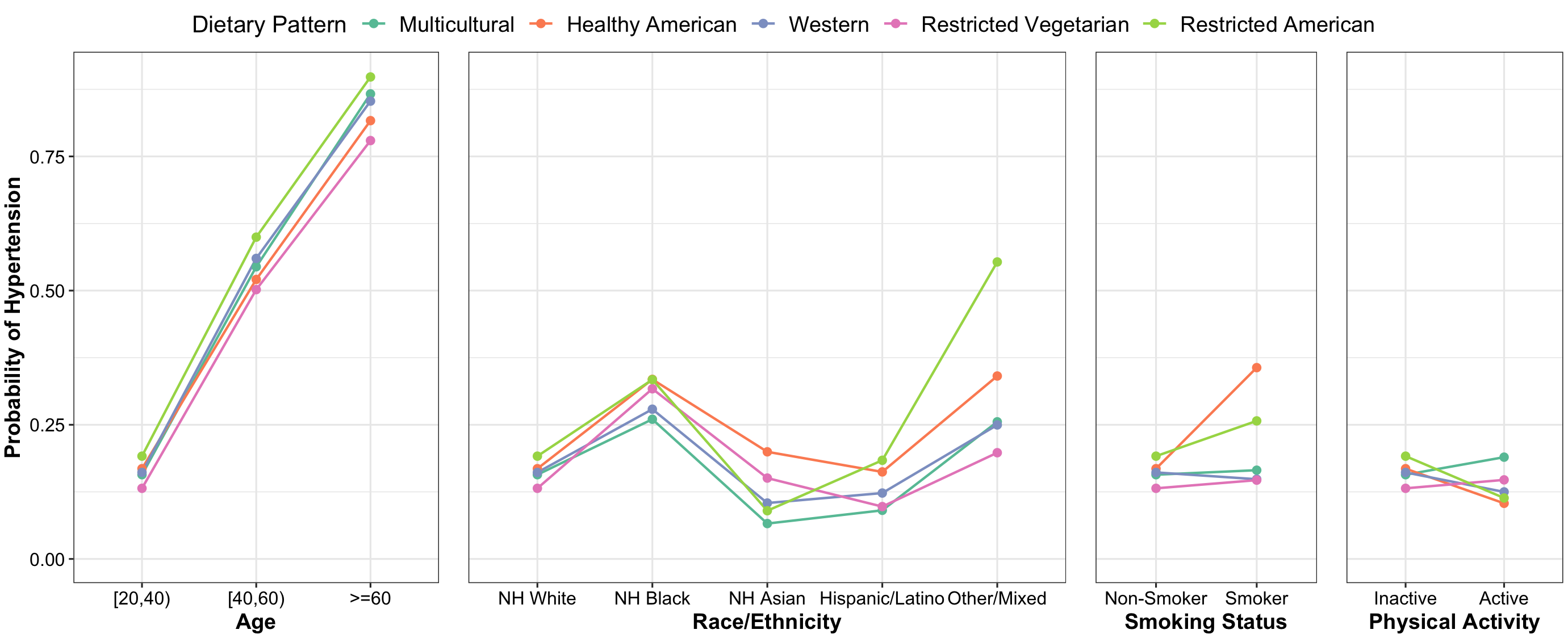}
    \label{f:SOLCA_phi}
\end{figure}

\subsubsection{Web Figure~\ref{f:SOLCA_phi}: SOLCA Hypertension Probability by Covariate}

\subsubsection{Web Table~\ref{t:SOLCA_coefs}: SOLCA Full Regression Output}
\begin{table}[H]
    \footnotesize
    \centering
    \caption{Full regression parameter estimates for unweighted SOLCA, adjusting for demographic confounders. \textmd{Reference group: Multicultural diet, age [20,40), NH White, non-smoker, physically inactive. Class2 = Healthy American, Class3 = Western, Class4 = Restricted Vegetarian, Class5 = Restricted American.}}
    \begin{tabular}{l>{}c>{}c>{}c}
    \toprule
    Covariate & Median & 95\% Credible Interval & P($\xi$ > 0)\\
    \midrule
    Intercept & -1.01 & (-1.61, -0.41) & <0.01\\
    Healthy Amer & 0.05 & (-0.68,  0.81) & 0.56\\
    Western & 0.02 & (-0.71,  0.70) & 0.53\\
    Restricted Veg & -0.11 & (-0.89,  0.64) & 0.39\\
    Restricted Amer & 0.14 & (-0.65,  0.91) & 0.63\\
    {}[40,60) & 1.12 & (0.67, 1.57) & 1.00\\
    >=60 & 2.12 & (1.60, 2.62) & 1.00\\
    NH Black & 0.36 & (-0.19,  0.91) & 0.90\\
    NH Asian & -0.50 & (-1.21,  0.16) & 0.07\\
    Hispanic/Latino & -0.33 & (-0.90,  0.24) & 0.13\\
    Other/Mixed & 0.35 & (-0.55,  1.31) & 0.79\\
    Smoker & 0.03 & (-0.43,  0.47) & 0.56\\
    Active & 0.13 & (-0.26,  0.51) & 0.73\\
    \addlinespace
    {}[40,60):Class2 & -0.10 & (-0.67,  0.48) & 0.37\\
    {}[40,60):Class3 & 0.03 & (-0.54,  0.59) & 0.54\\
    {}[40,60):Class4 & 0.02 & (-0.58,  0.59) & 0.52\\
    {}[40,60):Class5 & 0.01 & (-0.62,  0.63) & 0.51\\
    >=60:Class2 & -0.26 & (-0.95,  0.40) & 0.22\\
    >=60:Class3 & -0.09 & (-0.72,  0.55) & 0.39\\
    >=60:Class4 & -0.24 & (-0.88,  0.41) & 0.24\\
    >=60:Class5 & 0.00 & (-0.63,  0.71) & 0.50\\
    \addlinespace
    NH Black:Class2 & 0.17 & (-0.58,  0.91) & 0.68\\
    NH Black:Class3 & 0.03 & (-0.63,  0.72) & 0.53\\
    NH Black:Class4 & 0.29 & (-0.42,  0.93) & 0.79\\
    NH Black:Class5 & 0.08 & (-0.66,  0.79) & 0.59\\
    NH Asian:Class2 & 0.63 & (-0.51,  1.75) & 0.87\\
    NH Asian:Class3 & 0.21 & (-0.81,  1.23) & 0.66\\
    NH Asian:Class4 & 0.58 & (-0.24,  1.43) & 0.92\\
    NH Asian:Class5 & 0.03 & (-0.90,  1.03) & 0.52\\
    Hispanic/Latino:Class2 & 0.30 & (-0.41,  1.00) & 0.81\\
    Hispanic/Latino:Class3 & 0.14 & (-0.50,  0.83) & 0.68\\
    Hispanic/Latino:Class4 & 0.15 & (-0.52,  0.87) & 0.67\\
    Hispanic/Latino:Class5 & 0.30 & (-0.38,  0.97) & 0.80\\
    Other/Mixed:Class2 & 0.21 & (-1.09,  1.45) & 0.63\\
    Other/Mixed:Class3 & -0.05 & (-1.17,  1.07) & 0.46\\
    Other/Mixed:Class4 & -0.10 & (-1.27,  1.03) & 0.43\\
    Other/Mixed:Class5 & 0.64 & (-0.72,  2.08) & 0.82\\
    \addlinespace
    Smoker:Class2 & 0.56 & (-0.05,  1.20) & 0.96\\
    Smoker:Class3 & -0.09 & (-0.65,  0.49) & 0.38\\
    Smoker:Class4 & 0.02 & (-0.54,  0.65) & 0.54\\
    Smoker:Class5 & 0.19 & (-0.41,  0.79) & 0.74\\
    \addlinespace
    Active:Class2 & -0.43 & (-0.94,  0.11) & 0.06\\
    Active:Class3 & -0.29 & (-0.76,  0.19) & 0.11\\
    Active:Class4 & -0.06 & (-0.54,  0.45) & 0.41\\
    Active:Class5 & -0.46 & (-0.96,  0.03) & 0.04\\
    \bottomrule
    \end{tabular}
    \label{t:SOLCA_coefs}
\end{table}

\clearpage
\bibliographystyle{biom} 
\bibliography{aim1_biometrics}